\RequirePackage[2020-02-02]{latexrelease}
\documentclass[aps,prd,superscriptaddress,showpacs,preprint,amsmath,amssymb]{revtex4}
\usepackage{graphicx, bm}
\usepackage{float}
\usepackage{subcaption}
\usepackage[usenames]{color}
\newcommand{\crr}[1]{{\color{red} #1 }}
\newcommand{\cbb}[1]{{\color{blue} #1 }}

\begin{document}

\draft

\title{New physics search with the new gauge boson $Z'$  of the bestest little Higgs model at the muon collider}


\author{ A. Guti\'errez-Rodr\'{\i}guez\footnote{alexgu@fisica.uaz.edu.mx}}
\affiliation{\small Facultad de F\'{\i}sica, Universidad Aut\'onoma de Zacatecas\\
         Apartado Postal C-580, 98060 Zacatecas, M\'exico.\\}

\author{ E. Cruz-Albaro\footnote{elicruzalbaro88@gmail.com}}
\affiliation{\small Facultad de F\'{\i}sica, Universidad Aut\'onoma de Zacatecas\\
         Apartado Postal C-580, 98060 Zacatecas, M\'exico.\\}


\author{D. Espinosa-G\'omez \footnote{david.espinosa@umich.mx}
}
\affiliation{\small Facultad de Ciencias F\'{\i}sico Matem\'aticas, Universidad Michoacana de San Nicol\'as de Hidalgo\\
            Avenida Francisco, J. M\'ujica S/N, 58060, Morelia, Michoac\'an, M\'exico.\\}

\author{T. Cisneros-P\'erez \footnote{tzihue@gmail.com}
}
\affiliation{\small Unidad Acad\'emica de Ciencias Qu\'{\i}micas, Universidad Aut\'onoma de Zacatecas\\
         Apartado Postal C-585, 98060 Zacatecas, M\'exico.\\}

\author{ David A. P\'erez-Carlos\footnote{dperezcarlos@gmail.com}}
\affiliation{\small Facultad de F\'{\i}sica, Universidad Aut\'onoma de Zacatecas\\
         Apartado Postal C-580, 98060 Zacatecas, M\'exico.\\}

\date{\today}

\begin{abstract}

The Bestest Little Higgs Model (BLHM) has attracted increasing attention in recent years, mainly because it can explain the hierarchy problem without fine-tuning by introducing one-loop corrections to the Higgs boson mass through heavy top quark partners and heavy gauge bosons. 
In the context of this new model, we exhaustively investigated the impact of the BLHM parameters on the Higgs-strahlung production processes $\mu^{-} \mu^{+} \to Z'h_0$ and  $\mu^{-} \mu^{+} \to Z'H_0$, 
where $Z'$ and $H_0$ represent a new heavy gauge boson and a new heavy Higgs boson, respectively.
We consider the integrated luminosities of ${\cal L}=2000, 20 000$ ${\rm fb^{-1}}$ and the respective center-of-mass energies of  $\sqrt{s}=3000, 10 000$ GeV projected for a future muon collider to estimate the number of expected Higgs boson production events $h_0$ or $H_0$ in association with a new gauge boson $Z'$ at the muon collider.   In addition, we derive and present the Feynman rules for scalars ($h_0, H_0, \phi^{0}, \eta^{0}, \sigma, H^{\pm}, \phi^{\pm}, \eta^{\pm}, A_0$) and include the self-interactions of Higgs bosons.

\end{abstract}


\vspace{5mm}

\maketitle


\section{Introduction}

One of the main goals of current and future high-energy colliders is the search for physics beyond the Standard Model (SM). For high-energy experiments (such as the LHC~\cite{FCC:2018vvp,Benedikt:2022kan,ZurbanoFernandez:2020cco}, lepton colliders~\cite{ILC:2007bjz,Brunner:2022usy,MuonCollider:2022xlm,AlAli:2021let}, among others), a fundamental part of their research program is the study of the physics related to the Higgs boson within and beyond the SM (BSM).   An accurate description of the properties of the Higgs boson is crucial to understanding the mechanism of electroweak symmetry breaking and the BSM physics. 

Since the discovery of the Higgs boson (with properties compatible with those predicted by the SM)~\cite{ATLAS:2012yve,CMS:2012qbp}, many theoretical and experimental studies have been carried out in search of additional scalar particles or new gauge bosons. However, no new elementary particles other than the Higgs boson or the SM gauge bosons have been found to date.
Nevertheless, experimental searches continue as there are multiple theoretical motivations for the existence of new elementary particles. Apart from theoretical questions such as the hierarchy problem~\cite{Schmaltz:2002wx}, the origin of fermion families~\cite{Lemmon:2013qba}, the strong CP problem~\cite{tHooft:1976rip}, and the flavor puzzle in the SM~\cite{Weinberg:1977hb}, there are several unexplained observational facts such as the observed dark matter, the baryon asymmetry of the universe, the masses of neutrinos, and so on.
Many theories that aim to solve these problems predict the existence of new particles, such as dark photons, heavy Higgs bosons, heavy gauge bosons, right-handed neutrinos, axions, monopoles, or new interactions. 

Two common complementary routes are followed at the colliders to search for BSM physics. If the new physics appears at high masses, higher center-of-mass energy is necessary, while if the new physics involves small couplings to SM particles, maximizing collider luminosity will be indispensable. 
In particular, a muon collider~\cite{Accettura:2023ked,MuonCollider:2022nsa} with a given nominal energy and luminosity is much more efficient than a proton collider with comparable energy and luminosity. This is because muons, like protons, can collide with a center-of-mass energy of 10 TeV or more in a relatively compact ring without the fundamental limitations of synchrotron radiation.  However, being point-like particles, unlike protons, their nominal center-of-mass collision energy $E_{\text{cm}}$ is fully available to produce high-energy reactions probing length scales as short as $1/E_{\text{cm}}$. 
 Muon colliders can also probe the physics of electroweak interactions, including the nature of the Higgs sector responsible for breaking the electroweak symmetry.
 A future muon collider has the advantage of clean signatures and high-statistics samples of the Higgs boson, which could, in turn, probe extended models in a much more meaningful way than searches at the LHC.

In this paper, we research the Higgs-strahlung process where a Higgs boson, $h_0$ (SM Higgs) or $H_0$ (BLHM Higgs), is produced
in association with a new gauge boson $Z'$, which is predicted by the BLHM. This gauge boson emerges at the TeV scale and is also predicted in several extended models~\cite{Zwirner:1987kxa,Salvioni:2009mt,Perelstein:2005ka,Espinosa-Gomez:2023xrq}. 
In the context of the BLHM, we explore the phenomenology of production of the $\mu^{+}\mu^{-}\rightarrow (Z, Z') \to Z' h_0$ and $\mu^{+}\mu^{-}\rightarrow (Z, Z') \to Z' H_0$ processes where we consider resonant and non-resonant effects.
These processes are essential channels since they offer us the opportunity to study the contributions of the $Z Z' h_0$, $Z' Z' h_0$, $Z Z' H_0$, and $Z'Z' H_0$ couplings to the Higgs-strahlung production process at the future muon collider.
On the other hand, due to a muon collider's high collision energy capabilities, cross sections for the corresponding processes produced via vector boson fusion could become significantly more important than Higgs-strahlung processes, at least for SM particle production. We are currently working on this complementary project, and it will be addressed in further work.

The remainder of the paper is structured as follows: A  review of the BLHM is presented in Section~\ref{BLH}.
 In Section~\ref{zh0H0}, we find the scattering amplitudes and cross sections of the processes  $\mu^{+}\mu^{-}\to Z' h_0/ Z' H_0$.  Section~\ref{results} is devoted to our numerical results. 
  Finally, in Section~\ref{conclusions}, we present our conclusions.
Complementarily,   in Appendix~\ref{Rules}, we present the Feynman rules for the scalar particles of the BLHM and their self-interactions.

\section{A brief review of the BLHM} \label{BLH}

The BLHM~\cite{BLHM-2010,Kalyniak:2013eva,Godfrey:2012tf,Cisneros-Perez:2023foe,Cruz-Albaro:2024vjk,Cruz-Albaro:2023pah,Cruz-Albaro:2022lks,Cruz-Albaro:2022kty,Cisneros-Perez:2024efk,Cisneros-Perez:2024onx} is based on two independent nonlinear sigma models ($\Sigma$ and $\Delta$). When the field $\Sigma$ acquires a vacuum expectation value (VEV), the global $SO(6)_A\times SO(6)_B$ symmetry is broken to the diagonal group $SO(6)_V$ at the $f$ energy scale,  while when the second field $\Delta$  acquires a VEV, the global $SU(2)_C \times SU(2)_D$ symmetry is broken to the diagonal subgroup $SU(2)$ to the $F$ scale with $F> f$. In the first stage, 15 pseudo-Nambu-Goldstone bosons that are parametrized as

\begin{equation}\label{Sigma}
\Sigma=e^{i\Pi/f}  e^{2i\Pi_{h}/f}e^{i\Pi/f},
\end{equation}

\noindent where $\Pi$ and $\Pi_h$ are  given by

\begin{eqnarray}\label{Pi}
  \Pi=
 \left(
 \begin{array}{c c c}
  i(\phi_a T^{a}_L +\eta_{a} T_{R}^{a})_{4\times 4} & 0 & 0 \\
  0 & 0 &   i\sigma/\sqrt{2} \\
  0 &-i\sigma/\sqrt{2} &  0
\end{array}
 \right),  \hspace{1cm}
 \Pi_h=\frac{i}{\sqrt{2}}
 \left(
 \begin{array}{c c c}
  0_{4\times4} & h_1 & h_2 \\
  -h_{1}^{T} & 0 &   0 \\
  -h_{2}^{T} & 0 &  0
 \end{array}
 \right).
\end{eqnarray}

\noindent  In these expressions, $\phi_{a}$ and $\eta_{a}$ ($ a = 1,2,3 $) are real triplets, $h_{1}$ and $h_{2}$ are $\bf{4}'$s of $SO(4)$,
and $\sigma$ a real singlet. For Higgs fields, their explicit representation is $h_{i}^{T}=(h_{i1}, h_{i2}, h_{i3}, h_{i4})$ (see Eqs.~(\ref{h11})-(\ref{h24})),
while $T^{a}_{L, R}$ denote the generators of the group $SO(6)$ which are provided in Refs.~\cite{BLHM-2010,PhenomenologyBLH}.
 Regarding the second stage of symmetry breaking, the pseudo-Nambu-Goldstone bosons of the  $\Delta$ sigma field are parametrized as:

\begin{equation}\label{Delta}
\Delta=F e^{2i \Pi_d/F},\, \,\, \, \, \Pi_d=\chi_a \frac{\tau^{a}}{2} \ \ (a=1,2,3),
\end{equation}

\noindent
$\chi_a$ represents the Nambu-Goldstone fields, and the $\tau_a$ correspond to the Pauli matrices generators of the $SU(2)$  symmetry.

\subsection{Scalar sector}

The Higgs fields, $h_1$ and $h_2$, are used to construct a Higgs potential that undergoes spontaneous symmetry breaking. The basic form of this potential is as follows~\cite{BLHM-2010,Kalyniak:2013eva,Schmaltz:2008vd,PhenomenologyBLH,Erikson}

\begin{equation}\label{Vhiggs}
V_{Higgs}=\frac{1}{2}m_{1}^{2}h^{T}_{1}h_1 + \frac{1}{2}m_{2}^{2}h^{T}_{2}h_2 -B_\mu h^{T}_{1} h_2 + \frac{\lambda_{0}}{2} (h^{T}_{1}h_2)^{2},
\end{equation}

\noindent  where the components of the Higgs doublets ($h_1$, $h_2$), $B_{\mu}$ and $ \lambda_{0} $ are explicitly expressed as

\begin{eqnarray}
h_{11} &=& \text{cos}\ \alpha\, h_0 - \text{sin}\ \alpha\, H_0 +v\, \text{sin}\ \beta, \label{h11} \\
h_{21} &=&\text{sin}\ \alpha\, h_0 + \text{cos}\ \alpha\, H_0 +v\, \text{cos}\ \beta,\\
h_{12} &=& \text{cos}\ \beta\, A_0,\\
h_{22} &=& \text{sin}\ \beta\, A_0,\\
h_{13} &=& \frac{1}{\sqrt{2}}\left( \text{cos}\, \beta (H^{-}+H^{+}) \right),\\
h_{14} &=& \frac{i}{\sqrt{2}}\left( \text{cos}\, \beta (H^{-}-H^{+}) \right),\\
h_{23} &=& \frac{1}{\sqrt{2}}\left( \text{sin}\, \beta (H^{-}+H^{+}) \right),\\
h_{24} &=& \frac{i}{\sqrt{2}}\left( \text{sin}\, \beta (H^{-}- H^{+}) \right), \label{h24} \\
B_{\mu} &=& 2 \frac{\lambda_{56} m^{2}_{65} + \lambda_{65} m^{2}_{56} }{\lambda_{56} +\lambda_{65} }, \\
\lambda_{0} &=& 2 \frac{\lambda_{56} \lambda_{65} }{\lambda_{56} + \lambda_{65}}.
\end{eqnarray}

\noindent For the Higgs potential to reach a minimum, one must have $m_1 m_2 >0$, while electroweak symmetry breaking requires that $B_{\mu}>m_1 m_2$.

\noindent  Electroweak symmetry breaking in the BLHM is implemented when the Higgs doublets acquire their VEVs, $\langle h_1\rangle ^{T}=(v_1,0,0,0)$ and $ \langle h_2 \rangle ^{T}=(v_2,0,0,0)$. These VEVs minimize the Higgs potential of Eq.~(\ref{Vhiggs}), thus generating the relations:

\begin{eqnarray}\label{v12}
&&v^{2}_1=\frac{1}{\lambda_0}\frac{m_2}{m_1}(B_\mu-m_1 m_2),\\
&&v^{2}_2=\frac{1}{\lambda_0}\frac{m_1}{m_2}(B_\mu-m_1 m_2).
\end{eqnarray}

\noindent The VEVs can be expressed in terms of the parameters $v$ (the SM VEV) and $\tan \beta$ as follows

\begin{equation}\label{vvacio}
v^{2} \equiv v^{2}_1 +v^{2}_2= \frac{1}{\lambda_0}\left( \frac{m^{2}_1 + m^{2}_2}{m_1 m_2} \right) \left(B_\mu - m_1 m_2\right)\simeq \left(246\ \ \text{GeV}\right)^{2},
\end{equation}

\begin{equation}\label{beta}
\text{tan}\, \beta=\frac{v_1}{v_2}=\frac{m_2}{m_1}.
\end{equation}

\noindent After electroweak symmetry breaking, the scalar sector of the BLHM  generates several massive states:  two physical scalar fields ($H^{\pm}$)
and three neutral physical scalar fields ($h_0$, $H_0$, $A_0$). The lightest state, $h_0$, is identified as the Higgs boson of the SM. On the other hand, the four parameters in the Higgs potential $ m_1,  m_2, B_\mu$, and $\lambda_0$ can be replaced by another more phenomenologically accessible set~\cite{Kalyniak:2013eva}. That is, the masses of the states $h_0$ and $A_0$, the angle $\beta$ and the VEV $v$:

\begin{eqnarray}\label{parametros}
1 < &\text{tan} & \beta  <  \sqrt{ \frac{2+2 \sqrt{\big(1-\frac{m^{2}_{h_0} }{m^{2}_{A_0}} \big) \big(1-\frac{m^{2}_{h_0} }{4 \pi v^{2}}\big) } }{ \frac{m^{2}_{h_0}}{m^{2}_{A_0}} \big(1+ \frac{m^{2}_{A_0}- m^{2}_{h_0}}{4 \pi v^{2}}  \big) } -1 }, \label{cotabeta}\\
B_\mu &=&\frac{1}{2}(\lambda_0  v^{2} + m^{2}_{A_{0}}  )\, \text{sin}\, 2\beta,\\
\lambda_0 &=& \frac{m^{2}_{h_{0}}}{v^{2}}\Big(\frac{  m^{2}_{h_{0}}- m^{2}_{A_{0}} }{m^{2}_{h_{0}}-m^{2}_{A_{0}} \text{sin}^{2}\, 2\beta }\Big),\\
\text{tan}\, \alpha &=& \frac{ B_\mu \text{cot}\, 2\beta+ \sqrt{(B^{2}_\mu/\text{sin}^{2}\, 2\beta)-2\lambda_0 B_\mu v^{2} \text{sin}\, 2\beta+ \lambda^{2}_{0} v^{4}\text{sin}^{2}\, 2\beta  }  }{B_\mu -\lambda_0 v^{2} \text{sin}\, 2\beta},\label{alpha}   \\
m^{2}_{H^{\pm}} &=& m^{2}_{A_{0}}, \\
m^{2}_{H_{0}} &=& \frac{B_\mu}{\text{sin}\, 2\beta}+ \sqrt{\frac{B^{2}_{\mu}}{\text{sin}^{2}\, 2\beta} -2\lambda_0 B_\mu v^{2} \text{sin}\, 2\beta +\lambda^{2}_{0} v^{4} \text{sin}^{2}\, 2\beta  }, \label{mH0}\\
m^{2}_{\sigma}&=&2\lambda_0 f^{2} \text{K}_\sigma  \ \  \text{with} \ \ \ \  1< K_{\sigma}< \frac{16 \pi^{2}}{\lambda_0 (8\pi -\lambda_0)}. \label{masaescalar}
\end{eqnarray}

\subsection{Gauge sector}

The kinetic terms of the gauge fields in the BLHM are given as follows:

\begin{equation}\label{Lcinetico}
\mathcal{L}=\frac{f^{2}}{8} \text{Tr}(D_{\mu} \Sigma^{\dagger} D^{\mu} \Sigma) + \frac{F^{2}}{4} \text{Tr}(D_\mu \Delta^{\dagger} D^{\mu} \Delta),
\end{equation}

\noindent where the covariant derivatives are given by

\begin{eqnarray}\label{derivadasC}
D_{\mu}\Sigma&=&\partial_{\mu} \Sigma +i g_A A^{a}_{1\mu} T^{a}_L \Sigma- i g_B \Sigma A^{a}_{2\mu} T^{a}_L+ i g_{Y} B^{3}_{\mu}(T^{3}_{R}\Sigma-\Sigma T^{3}_{R}),\\
D_{\mu}\Delta&=&\partial_{\mu} \Delta +i g_A A^{a}_{1\mu} \frac{\tau^{a}}{2}  \Delta- i g_B \Delta A^{a}_{2\mu} \frac{\tau^{a}}{2}.
\end{eqnarray}

\noindent In these expressions, $T^{a}_{L}$ are the generators of the group $SO(6)_A$ corresponding to the subgroup $SU(2)_{LA}$, while $T^3_R$ represents
the third component of the $SO(6)_B$ generators corresponding to the $SU(2)_{LB} $ subgroup, these matrices are provided in Ref.~\cite{BLHM-2010}.
$g_A$ and $A^{a}_{1\mu}$ denote the gauge coupling and field associated with the gauge bosons of $SU(2)_{LA}$. $g_B$ and $A^{a}_{2\mu}$
represent the gauge coupling and the field associated with $SU(2)_{LB}$, while $g_Y$ and $B^{3}_{\mu}$ denote the hypercharge and the field.
When $\Sigma$ and $\Delta$ get their VEVs, the gauge fields $A^{a}_{1\mu}$ and $A^{a}_{2\mu}$ are mixed to form a massless triplet
$A^{a}_{0\mu}$ and a massive triplet $A^{a}_{H\mu}$,

\begin{equation}\label{AA}
A^{a}_{0\mu}=\text{cos}\, \theta_g A^{a}_{1\mu} + \text{sin}\, \theta_g A^{a}_{2\mu}, \hspace{5mm} A^{a}_{H\mu}= \text{sin}\, \theta_g A^{a}_{1\mu}- \text{cos}\, \theta_g A^{a}_{2\mu},
\end{equation}

\noindent with the mixing angle

\begin{equation}\label{gagb}
s_g\equiv \sin \theta_g=\frac{g_A}{\sqrt{g_{A}^{2}+g_{B}^{2}} },\ \ c_g \equiv \cos \theta_g=\frac{g_B}{\sqrt{g_{A}^{2}+g_{B}^{2}} },
\end{equation}

\noindent
which are related to the electroweak gauge coupling $g$ through

\begin{equation}\label{g}
g=\frac{g_A g_B}{\sqrt{g^{2}_A + g^{2}_B}}.
\end{equation}

\noindent  On the other hand, the weak mixing angle is defined as

\begin{eqnarray}\label{angulodebil}
s_W&&\equiv\sin \theta_W = \frac{g'}{\sqrt{g^2+ g'^{2} }}, \ \
c_W \equiv\cos \theta_W= \frac{g}{\sqrt{g^2+ g'^{2} }}.
\end{eqnarray}

In the BLHM, the masses of the new  gauge bosons ( $Z'$, y $ W^{\prime \pm} $) are also generated:

\begin{eqnarray}
m^{2}_{Z'}&=&m^{2}_{W'^{\pm}} +  \frac{g^2 s^{2}_W v^4}{16 c^{2}_W (f^2+F^2)} \left(s^{2}_g -c^{2}_g \right)^{2}, \label{mzprima} \\
m^{2}_{W'^{\pm}}&=& \frac{g^2}{4 c^{2}_{g} s^{2}_{g}} \left(f^2+F^2 \right)  - m^{2}_{W^{\pm}}. \label{mwprima}
\end{eqnarray}

\subsection{Yang-Mills sector}

In this sector, the interactions between gauge bosons are determined, and the corresponding lagrangian is

\begin{eqnarray}\label{YM}
\mathcal{L}=F_{1\mu \nu} F^{\mu \nu}_{1}  + F_{2\mu \nu} F^{\mu \nu}_{2},
\end{eqnarray}

\noindent where the tensors $ F^{\mu \nu}_{1,2}$ are given as

\begin{eqnarray}
F^{\mu \nu}_{1} &=& \partial^{\mu} A^{\alpha \nu}_{1} - \partial^{\nu} A^{\alpha \mu}_{1} + g_{A} \sum_{b} \sum_{c} \epsilon^{a b c} A^{b \mu}_{1} A^{c \nu}_{1}, \\
F^{\mu \nu}_{2} &=& \partial^{\mu} A^{\alpha \nu}_{2} - \partial^{\nu} A^{\alpha \mu}_{2} + g_{B} \sum_{b} \sum_{c} \epsilon^{a b c} A^{b \mu}_{2} A^{c \nu}_{2}.
\end{eqnarray}

\noindent
The indices $a$, $b$, and $c$ run over the three gauge fields~\cite{Martin:2012kqb}.

\subsection{Fermion sector} \label{subsecfermion}

The fermions of the BLHM have a $SO(6)$ representation. Thus, the Lagrangian that describes the interactions of the fermions is~\cite{BLHM-2010}

\begin{eqnarray}\label{Ltop}
\mathcal{L}_t &=& y_1 f Q^{T} S \Sigma S U^{c} + y_2 f Q'^{T} \Sigma U^{c} +y_3 f Q^{T} \Sigma U'^{c}_{5} +y_b f q_{3}^{T}(-2 i T^{2}_{R} \Sigma) U^{c}_{b} \nonumber \\
&+& \sum_{i=1,2} y_u f q^{T}_i \Sigma u^{c}_{i} + \sum_{i=1,2} y_{d} f q^{T}_{i}(-2i T^{2}_{R} \Sigma) d^{c}_i
+\sum_{i=1,2,3} y_e f l^{T}_i (-2i T^{2}_{R} \Sigma) e^{c}_i + \text{h.c.},
\end{eqnarray}

\noindent where $S$ is the $SO(6)$ matrix $ S = \text{diag} (1,1,1,1, -1, -1) $. The fermionic multiplets $ Q^{T} $, $Q'^{T} $, $ q_{3}^{T} $, $ U^{cT} $, $ U'^{cT} $, $ U_{b}^{cT} $, $ q^{T}_i$, $l^{T}_i$, $ u^{cT}_i$, $d^{cT}_i$, and $e^{cT}_i $  are arranged as follows~\cite{BLHM-2010,Martin:2012kqb,PhenomenologyBLH}

\begin{eqnarray}\label{camposf}
Q^{T}&=&\frac{1}{\sqrt{2}}\left( \left(-Q_{a_1} -Q_{b_2}\right), i\left(Q_{a_1} -Q_{b_2} \right),  \left(Q_{a_2} -Q_{b_1}\right), i\left(Q_{a_2} -Q_{b_1}\right), Q_{5},Q_{6} \right),\\
Q'^{T}&=&\frac{1}{\sqrt{2}} (-Q'_{a_1}, iQ'_{a_1},Q'_{a_2},iQ'_{a_2},0,0 ),\\
q_{3}^{T}&=& \frac{1}{\sqrt{2}} (-\bar{t}_L, i\bar{t}_L,\bar{b}_L,i\bar{b}_L,0,0 ),\\
U^{cT}&=& \frac{1}{\sqrt{2}} \left( (-U^{c}_{b_1} -U^{c}_{a_2}), i (U^{c}_{b_1} -U^{c}_{a_2}),  (U^{c}_{b_2} -U^{c}_{a_1}), i (U^{c}_{b_2} -U^{c}_{a_1}), U^{c}_{5},U^{c}_{6} \right),\\
U'^{cT}&=&(0,0,0,0,U'^{c}_5,0),\\
U_{b}^{cT}&=&(0,0,0,0,b^{c},0), \\
 q^{T}_i &=&\frac{1}{\sqrt{2}} (-\bar{u}_{iL}, i \bar{u}_{iL}, \bar{d}_{iL}, i \bar{d}_{iL},0,0),\\
 l^{T}_i &=&\frac{1}{\sqrt{2}} (-\bar{\nu}_{iL}, i \bar{\nu}_{iL}, \bar{e}_{iL}, i \bar{e}_{iL},0,0),\\
 u^{cT}_i &=&(0,0,0,0,u^{c}_i,0),\\
 d^{cT}_i &=&(0,0,0,0,d^{c}_i,0),\\
 e^{cT}_i &=&(0,0,0,0,e^{c}_i,0).
\end{eqnarray}

\noindent 
The superscript $c$ in some fermionic multiplets represents the right-handed helicity state of fermionic fields.
 Please note that the indices in $u^{c}_i$ and $d^{c}_i$ run over the first two generations of SM fermions, while $e^{c}_i$ runs over the three generations.   
 Concerning the Yukawa couplings $y_f$ ($f=u,d,e,b$), these are associated with the masses of the fermions as

\begin{eqnarray}
y_f &=& \frac{m_f}{v\sin\beta} \left(1-\frac{v^2}{3f^2} \right)^{-1/2}.
\end{eqnarray}

\noindent On the other hand, $y_1$, $y_2$ and $y_3$ represent the Yukawa couplings that generate two study scenarios,  $y_2 > y_3$  scenario and  $y_2 < y_3$ scenario~\cite{Cruz-Albaro:2022kty,Cruz-Albaro:2023pah}.
 For simplicity, the Yukawa couplings $y_i$ ($i=1,2,3$)  are assumed to be real~\cite{BLHM-2010,Godfrey:2012tf}.

Since the top quark loops provide the largest divergent corrections to the Higgs mass in the SM, the heavy quark sector in the BLHM scenario is the most crucial for solving the hierarchy problem. The new heavy quarks arising in the BLHM are $ T $, $ T_5 $, $ T_6 $, $ T^{2/3} $, $ T^{5/3} $, and  $ B $, whose associated masses are given as

\begin{eqnarray}
  m^{2}_T &=& (y^{2}_1 + y^{2}_2)f^2 + \frac{9 v^{2}_1 y^{2}_1  y^{2}_2  y^{2}_3 }{(y^{2}_1 + y^{2}_2) (y^{2}_2 - y^{2}_3)}, \label{MT} \\
  m^{2}_{T_5} &=& (y^{2}_1 + y^{2}_3)f^2 - \frac{9 v^{2}_1 y^{2}_1  y^{2}_2  y^{2}_3 }{(y^{2}_1 + y^{2}_3) (y^{2}_2 - y^{2}_3)},\label{MT5} \\
  m^{2}_{T_6} &=&m^{2}_{T^{2/3}_b}=m^{2}_{T^{5/3}_b} =y^{2}_1 f^2,\label{mT6} \\
  m^{2}_B & =&(y^{2}_1 + y^{2}_2)f^2,\label{mB}
\end{eqnarray}

\noindent where  $v_1=v\sin \beta$.
The mass terms for the new quarks, Eqs.~(\ref{MT})-(\ref{mB}), are calculated under the assumption that $y_{2} \neq y_{3}$, otherwise the masses of $T$ and $T_{5}$ are degenerate at lowest order~\cite{Godfrey:2012tf,PhenomenologyBLH}.

\subsection{Currents sector}

This sector determines the kinetic terms of fermions, as well as their interactions with gauge bosons~\cite{BLHM-2010,PhenomenologyBLH,Cruz-Albaro:2024vjk}:
\begin{eqnarray}\label{LbaseW}
 \mathcal{L} &=& \bar{Q} \bar{\tau}^{\mu} D_{\mu}Q + \bar{Q}' \bar{\tau}^{\mu} D_{\mu}Q'- U^{c\dagger} \tau^{\mu} D_{\mu}U^{c}-  U'^{c\dagger} \tau^{\mu} D_{\mu}U'^{c} -  U_{b}^{c\dagger} \tau^{\mu} D_{\mu}U_{b}^{c} +\sum_{i=1,2}  q^{\dagger}_i \tau^{\mu} D_{\mu} q_i   \nonumber \\
 &+& \sum_{i=1,2,3}  l^{\dagger}_i \tau^{\mu} D_{\mu} l_i
 - \sum_{i=1,2,3}  e_i^{c\dagger} \tau^{\mu} D_{\mu} e^{c}_i - \sum_{i=1,2}  u_{i}^{c\dagger} \tau^{\mu} D_{\mu} u^{c}_{i} - \sum_{i=1,2}  d_{i}^{c\dagger} \tau^{\mu} D_{\mu} d^{c}_i,
\end{eqnarray}

\noindent where $\tau^{\mu}$ and $\bar{\tau}^{\mu}$  are defined according to Ref.~\cite{Spremier}.


\section{ Higgs-strahlung processes $\mu^+ \mu^- \to (Z,Z') \to Z'h_0$ and $\mu^+ \mu^- \to (Z,Z') \to  Z' H_0$} \label{zh0H0}

\subsection{The decay width of the gauge boson $Z'$} \label{width}

 In the context of the BLHM a new gauge boson $Z'$ arises where its main decay channels of this neutral gauge boson are $Z' \to f_i \bar f_i$ ($f_i\equiv b, B, t, T, T_5, T_6,  T^{2/3}, T^{5/3}$),  $Z' \to W^+W^-$, $Z' \to Zh_0$, and $Z' \to ZH_0$. From these processes,  the total decay width $\Gamma_{Z'}$ of the $Z'$ boson is determined:

\begin{eqnarray} \label{gammaTot}
    \Gamma_{Z'} &=&\sum_{f_i} \Gamma_{f_i \bar f_i}+ \Gamma_{WW}+ \Gamma_{Zh_0}+ \Gamma_{ZH_0}.
\end{eqnarray}

\noindent The different partial decay widths of the $Z'$ boson involved in Eq.~(\ref{gammaTot})  have been calculated in Ref.~\cite{Martinez-Martinez:2024lez}. 
It is essential to mention that the total decay width of the gauge boson $Z'$ is an indispensable parameter for determining the cross-section of the Higgs-strahlung processes $\mu^+ \mu^- \to Z'h_0$ and $\mu^+ \mu^- \to   Z' H_0$.

\subsection{ Higgs-strahlung production $\mu^+\mu^{-} \to Z'h_0$ }

\begin{figure}[H]
\center
\subfloat[]{\includegraphics[width=6.30cm]{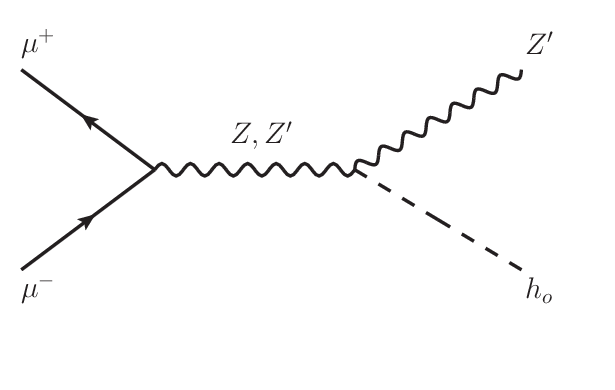}}
\subfloat[]{\includegraphics[width=6.25cm]{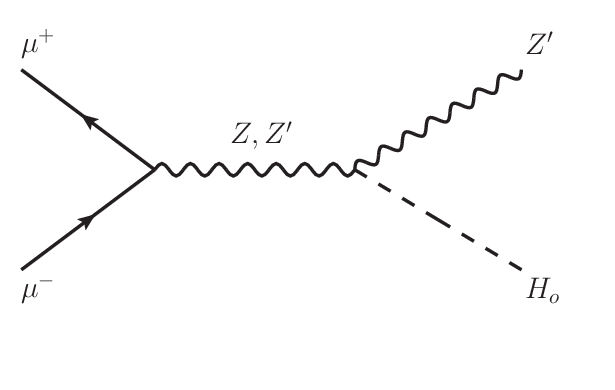}}
\caption{\label{strahlung}  Feynman diagrams  for the Higgs-strahlung production processes: a) $\mu^{+}\mu^{-}\to (Z,Z') \to Z'h_0$,   b)
 $\mu^{+}\mu^{-}\to (Z,Z') \to  Z'H_0$.}
\end{figure}

In Fig.~\ref{strahlung}(a), we show the Feynman diagram that represents the Higgs-strahlung production process  $\mu^+\mu^{-} \to (Z, Z') \to Z'h_0$.
From this Feynman diagram, we calculate the corresponding scattering amplitudes when the mediator particles are the gauge bosons $Z$ or $Z'$, such amplitudes are given in Eqs.~(\ref{MZ}) and~(\ref{MZp}).

\begin{eqnarray}
{\cal M}_Z (\mu^+\mu^{-}  \to Z'h_0) &=& g_{ZZ'h_0} \Bigl[ \bar v(p_1)\gamma^\mu (g^{Z\mu \mu}_V-g^{Z\mu \mu}_A\gamma_5)u(p_2) \Bigr ]
 \Bigl[ \frac{(-g_{\mu\nu} + p_{\mu}p_{\nu}/m^{2}_{Z})}{(p_{1}+p_{2})^{2}-m^{2}_{Z}-i m_{Z} \Gamma_{Z} }\Bigr] \nonumber \\
 &\times &  \epsilon^{*\nu}_\lambda(Z'),\label{MZ}\\
{\cal M}_{Z'} (\mu^+\mu^{-} \to Z'h_0) &=&  g_{Z^{'}Z^{'}h_0} \Bigl[ \bar v(p_1)\gamma^\mu (g^{Z\prime \mu \mu}_V-g^{Z\prime \mu \mu}_A\gamma_5)u(p_2) \Bigr ]
\Bigl[ \frac{(-g_{\mu\nu} + p_{\mu}p_{\nu}/m^{2}_{Z'})}{(p_{1}+p_{2})^{2}-m^{2}_{Z'}-i m_{Z^{\prime}} \Gamma_{Z'}}\Bigr]\nonumber \\
 &\times&  \epsilon^{*\nu}_\lambda(Z'), \label{MZp}
\end{eqnarray}

\noindent  where $\epsilon^{*\nu}_\lambda(Z')$  represents the polarization vector of the $Z'$ boson. 
On the other hand, $g^{Z\mu \mu}_{V}$ ($g^{Z' \mu \mu}_{V}$) and $g^{Z\mu \mu}_{A}$ ($g^{Z' \mu \mu}_{A}$) represent the vector and vector-axial coupling constants of the $Z$ (or $Z^{\prime})$ boson, while $g_{ZZ'h_0}$ and $g_{Z^{'}Z^{'}h_0}$ denote the effective couplings.

We calculate from the transition amplitudes the following cross-sections

\begin{eqnarray}
 \sigma^{Z'h_0}_{Z} &=& \frac{\lambda}{192\pi} \left(\frac{(g^{Z\mu \mu}_V)^2 + (g^{Z\mu \mu}_A)^2}{m_{Z'}^2}\right) \left(\frac{g_{ZZ'h_0}^2}{(s-m_{Z}^2)^2+(m_{Z}\Gamma_{Z})^2}\right) \left(12m_{Z'}^2 + \lambda^2\,s \right), \label{SZ} \\
 \sigma^{Z'h_0}_{Z'} &=& \frac{\lambda}{192\pi} \left(\frac{(g^{Z'\mu \mu}_V)^2 + (g^{Z'\mu \mu}_A)^2}{m_Z^2}\right) \left(\frac{g_{Z'Z'h_0}^2}{(s-m_{Z'}^2)^2+(m_{Z'}\Gamma_{Z'})^2}\right) \left(12m_{Z'}^2 + \lambda^2\,s \right), \label{SZp} \\
   \sigma^{Z'h_0}_{Z-Z'} &=& \lambda \left( \frac{g_{ZZ'h_0} g_{Z'Z'h_0}}{96\pi} \right) \left(\frac{g^{Z\mu \mu}_V g^{Z'\mu \mu}_{V} + g^{Z\mu \mu}_A g^{Z'\mu \mu}_A}{m_{Z'}^2} \right) \left(12m_{Z'}^2 + \lambda^2\,s \right)  \nonumber \\
     &\times& \frac{(s - m_Z^2)(s - m_{Z'}^2) + (m_Z \Gamma_Z)(\Gamma_{Z'}m_{Z'})}{((m_Z^2 - s)^2 + (m_Z\Gamma_Z)^2)((m_{Z'}^2 -
      s)^2 + (m_{Z'}\Gamma_{Z'})^2)}, \label{SZZp}
\end{eqnarray}

\noindent where $\sigma^{Z'h_0}_Z$ and  $\sigma^{Z'h_0}_{Z'}$  are the cross sections of processes $\mu^+ \mu^- \to Z \to Z'h_0$ and  $\mu^+ \mu^- \to Z' \to Z'h_{0}$, respectively. The $\sigma^{Z'h_0}_{Z-Z'}$ cross-section represents the interference effect between the new gauge boson $Z'$ and the SM gauge boson $Z$.
In Eqs. (\ref{SZ})-(\ref{SZZp}), $\sqrt{s}$ is the center-of-mass energy  and $\lambda$ is the two-particle phase space function,

\begin{equation}
    \lambda(s,m_{Z'},m_{h_0}) = \bigg[\bigg(1-\frac{m_{Z'}^2}{s}-\frac{m_{h_0}^2}{s}\bigg)^2 - \frac{4m_{Z'}^2 m_{h_0}^2}{s^2}\bigg]^{1/2}.
\end{equation}

As for the total cross section $ \sigma^{Z'h_0}_{T}$ for the processes $\mu^+ \mu^- \to (Z,Z') \to Z'h_0$, this is obtained as

\begin{equation}
\label{XSTotEq}
    \sigma^{Z'h_0}_{T}= \sigma^{Z'h_0}_Z + \sigma^{Z'h_0}_{Z'} + \sigma^{Z'h_0}_{Z-Z'}.
\end{equation}

\subsection{ Higgs-strahlung production $\mu^+\mu^{-} \to Z' H_0$ }

We determine the scattering amplitudes of the processes $\mu^+\mu^- \to Z \to Z' H_0$ and $\mu^+\mu^- \to  Z' \to Z'H_0$. From the Feynman diagram shown in Fig.~\ref{strahlung}(b), we derive the following amplitudes

\begin{eqnarray}
{\cal M}_Z (\mu^+\mu^{-}  \to Z'H_0) &=&  g_{ZZ'H_0} \Bigl[ \bar v(p_1)\gamma^\mu (g^{Z \mu \mu}_V-g^{Z \mu \mu}_A\gamma_5)u(p_2) \Bigr ]
\Bigl[ \frac{(-g_{\mu\nu} + p_{\mu}p_{\nu}/m^{2}_{Z})}{(p_{1}+p_{2})^{2}-m^{2}_{Z}-i m_{Z} \Gamma_{Z}} \Bigr] \nonumber \\
&\times &  \epsilon^{*\nu}_\lambda(Z'), \label{SZH0} \\
{\cal M}_{Z'} (\mu^+\mu^{-}  \to Z'H_0) &=&   g_{Z'Z'H_0} \Bigl[ \bar v(p_1)\gamma^\mu (g^{Z' \mu \mu}_V-g^{Z' \mu \mu}_A\gamma_5)u(p_2) \Bigr ]
\Bigl[\frac{(-g_{\mu\nu} + p_{\mu}p_{\nu}/m^{2}_{Z'})}{(p_{1}+p_{2})^{2}-m^{2}_{Z'}-i m_{Z'}\Gamma_{Z'}} \Bigr]  \nonumber \\
&\times & \epsilon^{*\nu}_\lambda(Z'). \label{SZPH0}
\end{eqnarray}

\noindent Using Eqs.~(\ref{SZH0}) and~(\ref{SZPH0}),  we calculate the cross sections

\begin{eqnarray}
   \sigma^{Z'H_0}_{Z} &=& \frac{\lambda}{192\pi} \left(\frac{(g^{Z\mu \mu}_V)^2 + (g^{Z\mu \mu}_A)^2 }{m_{Z'}^2}\right) \left(\frac{g_{ZZ'H_0}^2}{(s-m_{Z}^2)^2+(m_{Z}\Gamma_{Z})^2}\right) \left(12m_{Z'}^2 + \lambda^2\,s \right), \label{ZH0-Z}\\
\sigma^{Z'H_0}_{Z'} &=& \frac{\lambda}{192\pi} \left(\frac{(g^{Z' \mu \mu}_V)^2 + (g^{Z' \mu \mu}_A)^2}{m_{Z'}^2}\right) \left(\frac{g_{Z'Z'H_0}^2}{(s-m_{Z'}^2)^2+(m_{Z'}\Gamma_{Z'})^2}\right) \left(12m_{Z'}^2 + \lambda^2\,s \right), \label{ZH0-Zp} \\
 \sigma^{Z'H_0}_{Z-Z'} &=& \lambda \left( \frac{g_{Z'Z'H_0} g_{ZZ'H_0}}{96\pi} \right) \left(\frac{ g^{Z\mu \mu}_V g^{Z'\mu \mu}_{V} + g^{Z\mu \mu}_A g^{Z'\mu \mu}_A }{m_{Z'}^2} \right) \left(12m_{Z'}^2 + \lambda^2\,s\right)  \nonumber \\
     &\times& \frac{(s - m_Z^2)(s - m_{Z'}^2) + (m_Z \Gamma_Z)(\Gamma_{Z'}m_{Z'})}{((m_Z^2 - s)^2 + (m_Z\Gamma_Z)^2)((m_{Z'}^2 - s)^2 + (m_{Z'}\Gamma_{Z'})^2)}, \label{ZH0-mezcla}
\end{eqnarray}

\noindent where $\lambda$ is given by

\begin{equation}
   \lambda(s,m_{Z'},m_{H_0}) = \bigg[\bigg(1-\frac{m_{Z'}^2}{s}-\frac{m_{H_0}^2}{s}\bigg)^2 - \frac{4m_{Z'}^2 m_{H_0}^2}{s^2}\bigg]^{1/2}.
\end{equation}

The total $Z' H_0$ production cross section is derived as follows

\begin{equation} \label{ZH0tot}
    \sigma^{Z'H_0}_{T}= \sigma^{Z'H_0}_Z + \sigma^{Z'H_0}_{Z'} + \sigma^{Z'H_0}_{Z-Z'}.
\end{equation}

\section{Numerical results} \label{results}

For our numerical analysis of the contributions of the BLHM to the Higgs-strahlung processes $\mu^+\mu^- \to Z'h_0$ and $\mu^+\mu^- \to Z'H_0$. In Table~\ref{parametervalues}, we present the various theoretical and experimental constraints on the parameter space of the model of interest. Several LHC measurements are used to constrain the relevant parameters of the BLHM.

With respect to the cross-section of the  $\mu^+\mu^- \to Z'h_0$ and $\mu^+\mu^- \to Z'H_0$ processes, as well as the decay width $\Gamma_{Z'}$, these observables offer several possible scenarios of study, which is because the interaction vertices involved such as $ZZ'h_0$, $ZZ'H_0$, $Z'Z'h_0$, and $Z'Z'H_0$ depend on the gauge couplings $g_A$ and $g_B$. We have selected the following scenarios for our analysis in the present study: $g_B=\frac{1}{2}g_A$ and $g_B=g_A$. With respect to the selected scenarios, from Eqs.~(\ref{gagb}) and~(\ref{g}), we derive the values $g_A= \sqrt{5}\, g $ and $ g_A=\sqrt{2}\, g $, respectively.

\begin{table}[H]
\caption{Parameters involved in our numerical analysis of the $\mu^+\mu^{-} \to Z' h_0$ and $\mu^+\mu^{-} \to Z' H_0$ processes.
\label{parametervalues}}
\centering
\begin{tabular}{|c | c | c |}
\hline
\hspace{0.5cm} $ \textbf{Parameter} $ \hspace{0.5cm}  &  \hspace{1.2cm}  $\textbf{Value} $ \hspace{1.2cm}  &  \hspace{0.5cm}   $ \textbf{Reference} $ \hspace{0.5cm} \\
\hline
\hline
\hline
$ g_{A} $ (\text{when} $g_B= \frac{1}{2}g_A$)  &   $ \sqrt{5}\, g $ & \cite{Martinez-Martinez:2024lez} \\
\hline
$ g_{A} $  (\text{when}  $g_B=g_A$) &   $ \sqrt{2}\, g $ & \cite{Cruz-Albaro:2024vjk,Cruz-Albaro:2023pah,Cruz-Albaro:2022kty,Cruz-Albaro:2022lks} \\
\hline
$ m_{h_{0}}  $  &   $ 125.25\  \text{GeV} $ &  \cite{Workman:2022ynf}  \\
\hline
$ m_{A_{0}}  $  &   $ 1000\  \text{GeV} $ &   \cite{ATLAS:2020gxx,CMS:2019ogx}  \\
\hline
$ \tan \beta $  &    $ (1,10.45) $  &  \cite{Cruz-Albaro:2023pah,Cruz-Albaro:2022kty,Cruz-Albaro:2022lks} \\
\hline
$ y_{1} $   &   $ 0.61 $ & \cite{Cruz-Albaro:2023pah,Cruz-Albaro:2022lks} \\
\hline
$ y_{2} $   &   $ 0.35 $ & \cite{Cruz-Albaro:2023pah,Cruz-Albaro:2022lks} \\
\hline
$ y_{3} $   &   $ 0.84 $ & \cite{Cruz-Albaro:2023pah,Cruz-Albaro:2022lks} \\
\hline
$\Gamma_Z$  &  $2.4952\pm 0.0023$ GeV  & \cite{Workman:2022ynf} \\
\hline
$ f $  &  $ [1000, 2000]\   \text{GeV} $ &  \cite{BLHM-2010,Godfrey:2012tf,Cruz-Albaro:2023pah, Cruz-Albaro:2022kty,Cruz-Albaro:2022lks}  \\
\hline
$ F $  &   $ > 3000  \ \text{GeV} $ &  \cite{BLHM-2010,Kalyniak:2013eva,Godfrey:2012tf,Cruz-Albaro:2023pah,Cruz-Albaro:2022kty,Cruz-Albaro:2022lks} \\
\hline
\end{tabular}
\end{table}

Other important parameters in our calculations are the following:

 \noindent \textbf{The mass of the Higgs boson  $H_{0}$}: 
 The Higgs boson mass $H_{0}$ is determined from Eq.~(\ref{mH0}) and depends on the parameters $m_{A_0}$ and $\tan \beta$.
\vspace{0.1cm}

 \noindent \textbf{The mass of the gauge boson $Z'$}: In the BLHM scenario, the neutral gauge boson obtains large masses proportional to $\sqrt{f^{2}+F^{2}}$ (see Eq.~(\ref{mzprima})).
\vspace{0.1cm}

\subsection{ Production of the Higgs bosons $h_0$, $H_0$ and the gauge boson $Z'$ under the $g_{B}=1/2 \, g_{A}$ scenario}

\subsubsection{$\Gamma_{Z^{\prime}} $}

\begin{figure}[H]
\center
\subfloat[]{\includegraphics[width=8.30cm]{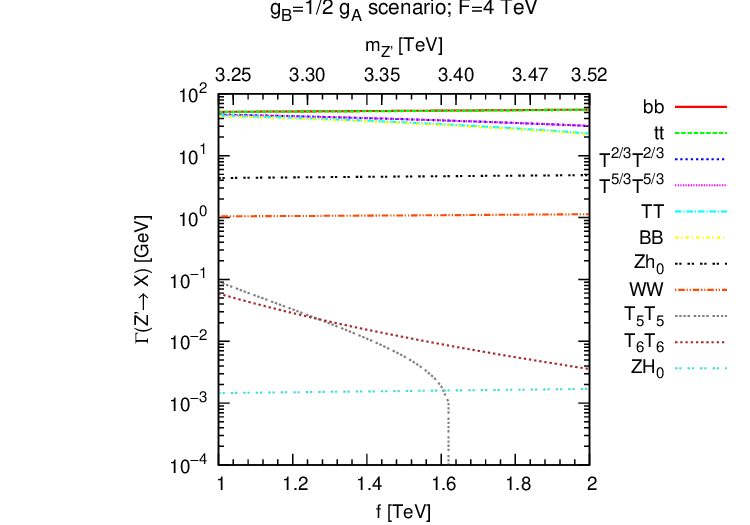}}
\subfloat[]{\includegraphics[width=8.30cm]{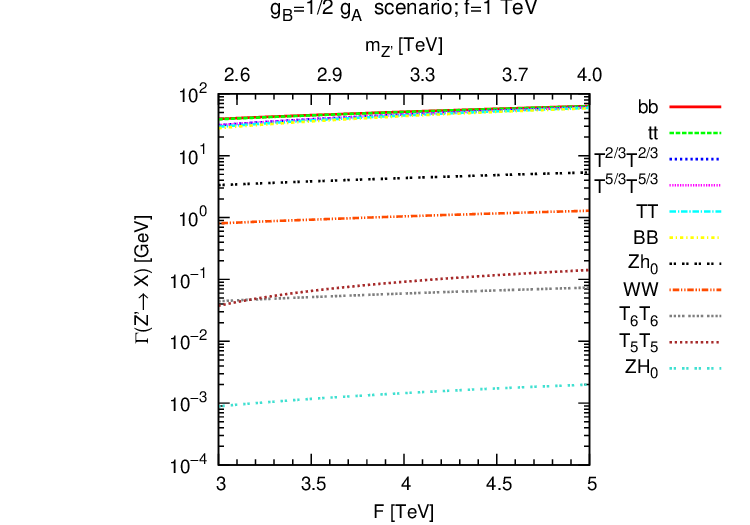}}
\caption{ \label{widths} Decay widths for the processes  $Z' \to X$  where $X=b\bar b, B\bar B, t\bar t,  T\bar T, T_5\bar T_5,$ $ T_6\bar T_6, T^{2/3}\bar T^{2/3}, T^{5/3}\bar T^{5/3},  Z h_{0}, Z H_{0}, W W$. a) $\Gamma(Z^{\prime} \to X)$ as a function of the $f$ energy scale (with  $F=4\, 000$ GeV). b) $\Gamma(Z^{\prime}  \to X)$ as a function of the $F$ energy scale (with $f=1\, 000$ GeV). }
\end{figure}

A new gauge boson $Z'$  is predicted in the framework of the BLHM where its main decay modes in the   $g_B=\frac{1}{2} g_A$ scenario are  $b\bar b, B\bar B, t\bar t,  T\bar T, T_5\bar T_5$, $ T_6\bar T_6, T^{2/3}\bar T^{2/3}, T^{5/3}\bar T^{5/3},  Z h_{0}, Z H_{0}$, and $W W$.
For these processes, their corresponding decay widths are determined. Fig.~\ref{widths} shows the behavior of the generated curves representing the decay widths $\Gamma(Z' \to X)$ as a function of the scale $f$ or $F$ where the established analysis intervals are $[1000,2000]$ GeV and $[3000, 5000]$ GeV, respectively. When we set $F =4000$ GeV, we can appreciate the evolution of $\Gamma(Z' \to X)$ vs. $f$ in Fig.~\ref{widths}(a). 
In this context, the curves that provide the dominant and subdominant contributions are given by $\Gamma(Z' \to b\bar b)$ and $\Gamma(Z' \to  t\bar t)$: $\Gamma(Z' \to b\bar b)=[51.30,55.64]$ GeV and $\Gamma(Z' \to  t\bar t)=[50.88,55.22]$ GeV.
Complementarily, the curve generating the suppressed contribution is provided by  $\Gamma(Z' \to  Z H_0)$ while $f\in [1000, 1620]$ GeV, for this interval $\Gamma(Z' \to  Z H_0) \in [1.45,1.60]\times 10^{-3}$ GeV.
On the other hand, when we fixed the $f$ scale to 1000 GeV, we generated Fig.~\ref{widths}(b), which shows the dependence of $\Gamma(Z' \to X)$ on $F$.
 As in the previous case, the decay channels $Z' \to b\bar b $ and $Z' \to  t\bar t$ provide the dominant and subdominant decay widths, $\Gamma(Z' \to b\bar b)=[39.35,63.44]$ GeV and $\Gamma(Z' \to  t\bar t)=[38.79,63.11]$ GeV. 
  As for the $Z' \to Z H_0 $ process, it generates the smallest decay width in the whole study range of the $F$ energy scale,   $\Gamma(Z' \to Z H_0)=[8.87 \times 10^{-4}, 2.00 \times 10^{-3}]$ GeV. 
  From Figs.~\ref{widths}(a) and~\ref{widths}(b), it can be observed that the fermionic decays of the $Z'$ gauge boson provide the most significant contributions to the decay width of the $Z'$ boson, except the $Z'\to  T_5\bar T_5$ and $Z' \to T_6\bar T_6$ processes which generate small contributions.
We also find that $\Gamma(Z' \to X)$ shows a higher sensitivity to the $F$ parameter changes than the $f$ scale. This same behavior exhibits $m_{Z'}$ (the mass of the gauge boson $Z'$) when varied versus parameter $F$, as can be seen on the upper horizontal axis of Fig.~\ref{widths}(b).

\subsubsection{$\text{Br}\left(Z^{\prime} \to X \right) $}

\begin{figure}[H]
\center
\subfloat[]{\includegraphics[width=8.20cm]{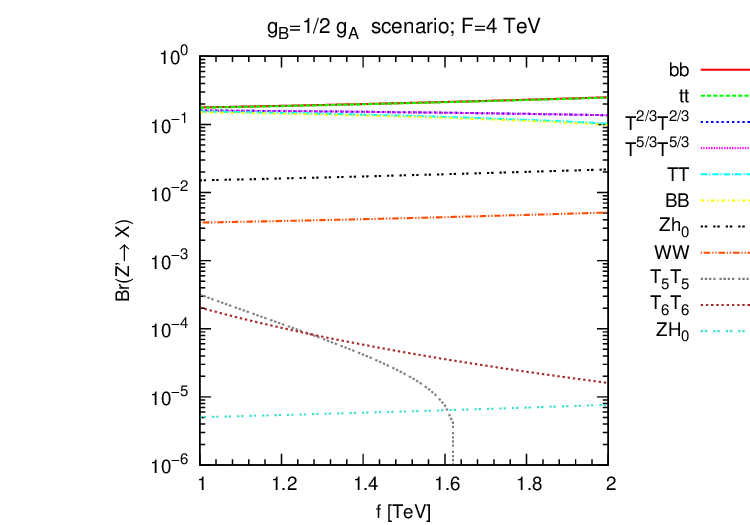}}
\subfloat[]{\includegraphics[width=8.20cm]{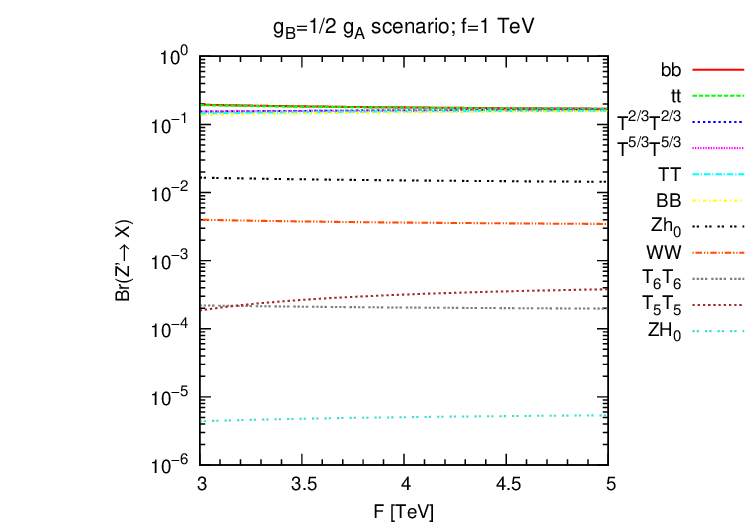}}
\caption{ \label{br} Branching ratios for the processes $Z' \to X$  where $X= b\bar b, B\bar B, t\bar t,  T\bar T, T_5\bar T_5,$ $ T_6\bar T_6, T^{2/3}\bar T^{2/3}, T^{5/3}\bar T^{5/3},  Z h_{0}, Z H_{0}, W W$. a) $\text{Br}(Z^{\prime} \to X)$ as a function of the $f$ energy scale (with  $F=4\, 000$ GeV). b) $\text{Br}(Z^{\prime}  \to X)$ as a function of the $F$ energy scale (with  $f=1\, 000$ GeV).
}
\end{figure}

 We calculate the branching ratios of the $Z' \to X$ processes as a function of the energy scale $f$ or $F$, as seen in Fig.~\ref{br}.
In the left plot of Fig.~\ref{br}, we analyze the dependence of $\text{Br}(Z' \to X)$ vs. $f$ and find that the main contributions are given by the $Z'\to b\bar b$ and $Z' \to t\bar t$ decays. Numerical evaluations generate $\text{Br}(Z' \to b\bar b)=[1.78,2.50]\times 10^{-1}$ and $\text{Br}(Z' \to t\bar t)=[1.76,2.49]\times 10^{-1}$ when $f\in [1000,2000]$ GeV. In contrast, the smallest contribution comes from the $Z'\to Z H_0$ decay while $f\in [1000,1620]$ GeV, $\text{Br}(Z' \to Z H_0)=[5.03,6.42]\times 10^{-6}$.
Other branching ratios that provide considerable contributions are $\text{Br}(Z' \to T^{2/3}\bar T^{2/3})=[1.60,1.36]\times 10^{-1}$,  $\text{Br}(Z' \to T^{5/3}\bar T^{5/3})=[1.60,1.37]\times 10^{-1}$,  $\text{Br}(Z' \to T\bar T)=[1.55,1.03] \times 10^{-1}$, $\text{Br}(Z' \to B\bar B)=[1.51 \times 10^{-1},9.85 \times 10^{-2}]$, and  $\text{Br}(Z' \to Z h_0)=[1.51,2.19] \times 10^{-2}$.
 With respect to the right plot in Fig.~\ref{br}, we show the values of  $\text{Br}(Z' \to X)$ when the $F$ scale acquires values from 3000 to 5000 GeV.
In this analysis interval, we obtain that the significant contributions to $\text{Br}(Z' \to X)$ are again generated by the tree-level decays $Z'\to b\bar b$ and $Z' \to t\bar t$: $\text{Br}(Z' \to b\bar b)=[1.95,1.70]\times 10^{-1}$ and $\text{Br}(Z' \to t\bar t)=[1.92,1.69]\times 10^{-1}$.
Additional processes that provide significant contributions to $\text{Br}(Z' \to X)$ are $Z' \to T^{2/3}\bar T^{2/3}$, $Z' \to T^{5/3}\bar T^{5/3}$,  $Z' \to T\bar T$, $Z' \to B\bar B$, and $Z' \to Z h_0$ whose corresponding branching ratios are as follows: 
$\text{Br}(Z' \to T^{2/3}\bar T^{2/3})\sim \text{Br}(Z' \to T^{5/3}\bar T^{5/3})=[1.54, 1.63]\times 10^{-1}$,  $\text{Br}(Z' \to T\bar T)=[1.45, 1.59] \times 10^{-1}$, $\text{Br}(Z' \to B\bar B)=[1.38, 1.57] \times 10^{-1}$, and $\text{Br}(Z' \to Z h_0)=[1.66, 1.44] \times 10^{-2}$.
As for the curve generating a suppressed contribution, it is given by $\text{Br}(Z' \to Z H_0)=[4.39, 5.36] \times 10^{-6}$.
In summary, $\text{Br}(Z' \to X)$ shows a dependence on the two energy scales, $f$ and $F$.

\subsubsection{Production of the Higgs boson $h_0$ and the new gauge boson $Z'$} \label{production-zp-h0}

\begin{figure}[H] 
\center
{\includegraphics[width=10.0cm]{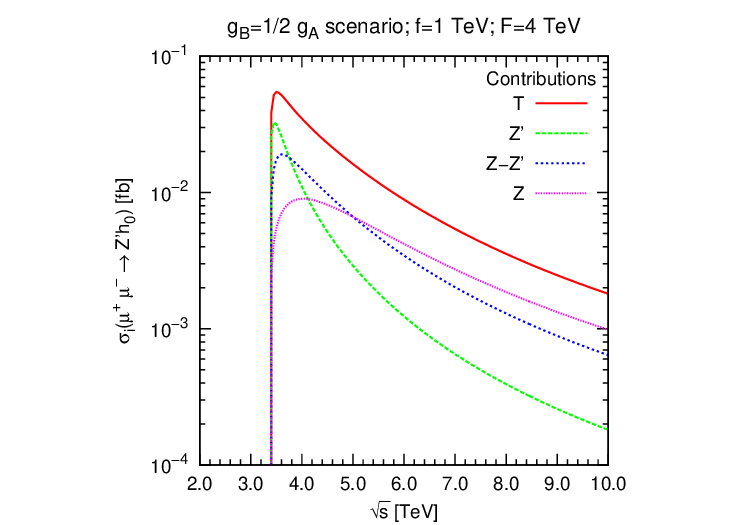}}
\caption{The different contributions to the cross section of the process  $\mu^{+}\mu^{-}\to (Z, Z') \to Z' h_0$   as a function of $\sqrt{s}$.} 
\label{Zph0}
\end{figure}

In this Subsection, we study the production of the SM Higgs boson $h_0$ in association with a new neutral gauge boson $Z'$ predicted in the context of the BLHM. We consider the contributions of the $ZZ'h_0$ and $Z'Z'h_0$ couplings to the Higgs-strahlung production process $\mu^{+} \mu^{-} \to Z' h_0$ at the future muon collider. To determine the contribution of the BLHM to the total $Z' h_0$ production cross section, the cross sections $\sigma^{Z' h_0}_{Z}$ and $\sigma^{Z' h_0}_{Z'}$ are estimated for the $\mu^{+} \mu^{-}\to Z \to Z' h_0$ and $\mu^{+} \mu^{-} \to Z' \to Z' h_0$ processes, as well as $\sigma^{Z' h_0}_{Z-Z'}$ which quantifies the interference effect between the $Z$ and $Z'$ gauge bosons. 
From $\sigma^{Z' h_0}_{Z}$, $\sigma^{Z' h_0}_{Z'}$, and $\sigma^{Z' h_0}_{Z-Z'}$ we compute the total cross section $\sigma^{Z' h_0}_{T}\left(\sqrt{s}, f, F, \beta \right)$ (see Eq.~(\ref{XSTotEq})). 
In Fig.~\ref{Zph0}, we show the contributions generated by each of the cross sections when we analyze the dependence of these versus the center-of-mass energy $\sqrt{s}$ which varies from 2000 to 10000 GeV, while the other parameters are set to $\tan \beta=3$, $f=1000$ GeV, and $F=4000$ GeV.
This figure shows that the curves involved reach their maximum peaks in the center-of-mass energy region from 3400 to 4100 GeV.
  Inside this region, the new gauge boson $Z'$  provides the largest contribution to $\sigma_{T}\left(\mu^{+} \mu^{-} \to Z' h_0 \right)$. The maximum peaks for each curve are $\sigma_{Z'}\left(\mu^{+} \mu^{-} \to Z' h_0 \right)=3.28 \times 10^{-2}$ fb ($\sqrt{s} \approx 3460$ GeV), $\sigma_{Z-Z'}\left(\mu^{+} \mu^{-} \to Z' h_0 \right)=1.91 \times 10^{-2}$ fb ($\sqrt{s} \approx 3610$ GeV), $\sigma_{Z}\left(\mu^{+} \mu^{-} \to Z' h_0 \right)=9.00 \times 10^{-3}$ fb ($\sqrt{s} \approx 4030$ GeV), and $\sigma_{T}\left(\mu^{+} \mu^{-} \to Z' h_0 \right)=5.48 \times 10^{-2}$ fb ($\sqrt{s} \approx 3510$ GeV).
The corresponding graph clearly shows that for the region $\sqrt{s}>4100$ GeV, $\sigma_{i}\left(\mu^{+} \mu^{-} \to Z' h_0 \right)$ ($i\equiv Z,Z',Z-Z',T$) decreases by about one or two orders of magnitude as $\sqrt{s}$  increases up to 10000 GeV. The curves demonstrate that the $\sqrt{s}$ parameter significantly impacts the results.

\begin{figure}[H]
\center
\subfloat[]{\includegraphics[width=8.20cm]{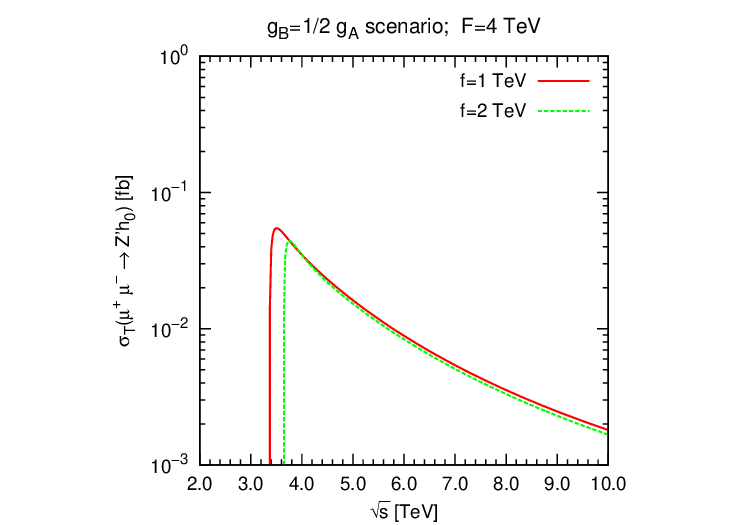}}
\subfloat[]{\includegraphics[width=8.20cm]{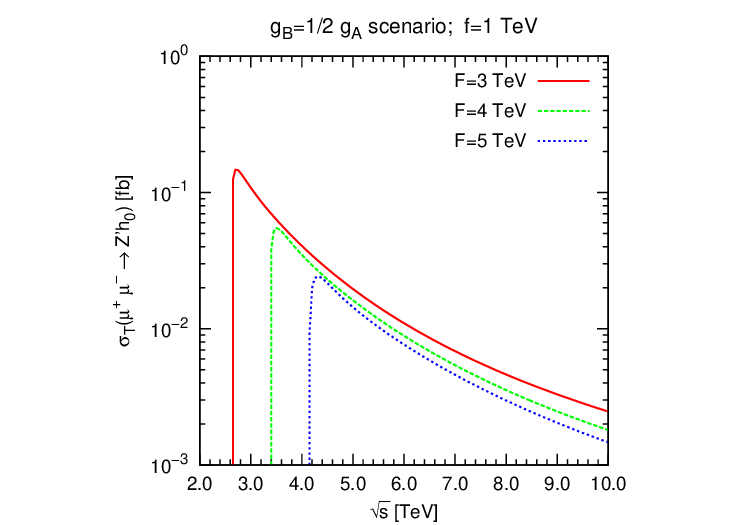}}
        \caption{The total cross section of the process $\mu^{+}\mu^{-}\to (Z, Z') \to Z' h_0$   as a function of $\sqrt{s}$.
        a)  The curves are generated for $f=1000$ GeV  and $f=2 000$ GeV.
     b)   The curves are generated for $F=3 000$ GeV,  $F=4000$ GeV, and  $F=5 000$ GeV.}
     \label{sfF}
\end{figure}

To see the effects of varying the values of the free parameters $\sqrt{s}$ and $f$ on the $Z'h_0$ production, we fix the value of the $F$ scale to 4000 GeV while $\tan \beta=3$.
 In this way, in Fig.~\ref{sfF}(a) we show the different curves generated for $\sigma_{T}\left(\mu^{+} \mu^{-} \to Z' h_0 \right)$  when the  $f$ scale takes specific fixed values while the center-of-mass energy $\sqrt{s}$ varies in the interval from 2000 to 10000 GeV. 
For smaller values of the $f$ scale larger values are generated on $\sigma_{T}\left(\mu^{+} \mu^{-} \to Z' h_0 \right)$, in particular, when $f=1000$ GeV the largest peak of the associated curve is reached:  $\sigma_{T}\left(\mu^{+} \mu^{-} \to Z' h_0 \right)=5.48\times 10^{-2}$ fb for $\sqrt{s} \approx 3510$ GeV. 
Complementarily, we now verify the effects that $\sqrt{s}$ and $F$ could provide on $\sigma_{T}\left(\mu^{+} \mu^{-} \to Z' h_0 \right)$ while we set the $f$ scale  to 1000 GeV.  As shown in Fig.~\ref{sfF}(b), the $Z'h_0$ production becomes smaller when the $F$ scale obtains larger values. The three curves are generated for $F=3000$ GeV, $F=4000$ GeV, and $F=5000$ GeV. The maximum peak on $\sigma_{T}\left(\mu^{+} \mu^{-} \to Z' h_0 \right)$ is found when $F=3000$ GeV: $\sigma_{T}\left(\mu^{+} \mu^{-} \to Z' h_0 \right)=1.48 \times 10^{-1}$ fb when $\sqrt{s} \approx 2720$ GeV.
From the figures discussed above, we can also see that the curves corresponding to $\sigma_{T}\left(\mu^{+} \mu^{-} \to Z' h_0 \right)$ decrease for large values of $\sqrt{s}$.

\begin{figure}[H] 
\center
{\includegraphics[width=10.cm]{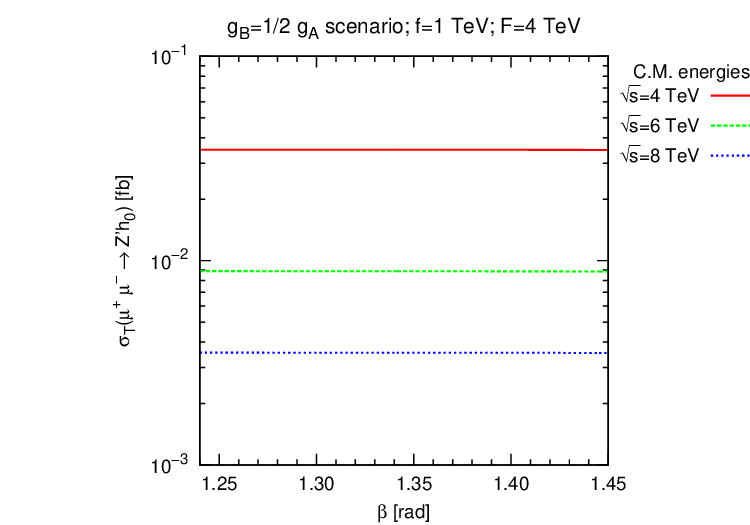}}
 \caption{The total cross section of the process $\mu^{+}\mu^{-}\to (Z, Z') \to Z' h_0$   as a function of parameter $\beta$ ($\tan \beta \in [3,9]$).
 The curves are generated for different values of the center-of-mass energy $\sqrt{s}$: $\sqrt{s}=4 000$ GeV, $\sqrt{s}=6 000$ GeV, and $\sqrt{s}=8 000$ GeV.}
 \label{shvsbeta}
\end{figure}

Below we study the behavior of $\sigma_{T}\left(\mu^{+} \mu^{-} \to Z' h_0 \right)$ as a function of $\beta$ angle  for center-of-mass energies $\sqrt{s}=4000, 6000, 8000$ GeV. The region of analysis of  $\beta$ angle  is set according to Eq.~(\ref{cotabeta}), so we have consistently chosen the range ($\tan^{-1} (3)$, $\tan^{-1} (9)$) as our study interval for $\beta$ parameter.
In Fig.~\ref{shvsbeta},  we can appreciate that the  $Z'h_0$ production is more abundant when $\sqrt{s}=4000$ GeV while for $\sqrt{s}=8000$ GeV a smaller production is obtained, $\sigma_{T}\left(\mu^{+} \mu^{-} \to Z' h_0 \right)=[3.49,3.48]\times 10^{-2}$ fb and $\sigma_{T}\left(\mu^{+} \mu^{-} \to Z' h_0 \right)=[3.55,3.53]\times 10^{-3}$ fb, respectively.
According to our numerical analysis,  we have found that the curves generated from the total cross section  $\sigma_{T}\left(\mu^{+} \mu^{-} \to Z' h_0 \right)$  show little sensitivity to changes in the values of $\beta$ parameter as far as they lie between the allowed intervals. On the contrary,  $\sigma_{T}\left(\mu^{+} \mu^{-} \to Z' h_0 \right)$  does show a stronger dependence on the $\sqrt{s}$ parameter.
It is essential to mention that the curves in Fig.~\ref{shvsbeta} have been generated for fixed values of the energy scales, $f=1000$ GeV and $F=4000$ GeV. However, the same behavior is obtained for any choice of the new physics scales.


\begin{figure}[H]
\center
\subfloat[]{\includegraphics[width=8.20cm]{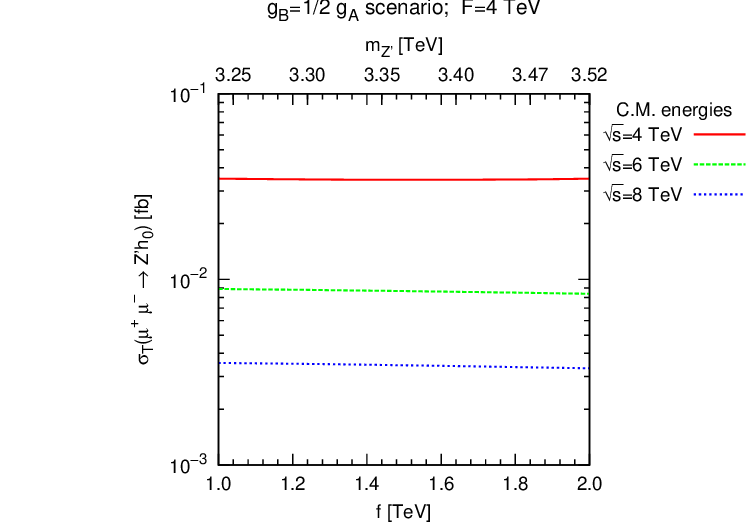}}
\subfloat[]{\includegraphics[width=8.20cm]{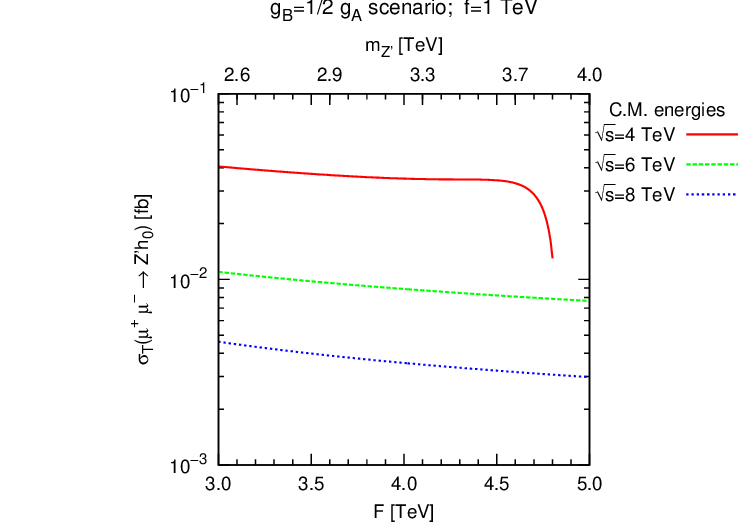}}
        \caption{ a) The total cross section of the process $\mu^{+}\mu^{-}\to (Z, Z') \to Z' h_0$   as a function of the $f$ energy scale.
        b) The total cross section of the process $\mu^{+}\mu^{-}\to (Z, Z') \to Z' h_0$   as a function of the $F$ energy scale.
        The curves are generated for different values of the center-of-mass energy $\sqrt{s}$: $\sqrt{s}=4000$ GeV, $\sqrt{s}=6000$ GeV, and $\sqrt{s}=8 000$ GeV.}
\label{fF-s}
\end{figure}

It is also convenient to analyze the behavior of $\sigma_{T}\left(\mu^{+} \mu^{-} \to Z' h_0 \right)$ as a function of the symmetry breaking scale $f
$ or $F$ since the masses of the new particles, specifically, the mass of the new gauge boson $Z'$ depend on them.
In this way, in Figs.~\ref{fF-s}(a) and~\ref{fF-s}(b) we can observe the curves generated for $\sigma_{T}\left(\mu^{+} \mu^{-} \to Z' h_0 \right)$ when we assign certain fixed values to the center-of-mass energy of the muon collider, $\sqrt{s}=4000, 6000, 8000$ GeV. 
On the upper horizontal axis, we also show the values of the mass of the new gauge boson $Z'$  when the scale of the new physics $f$ or $F$ takes values in the permissible range.
When we explore  $\sigma_{T}\left(\mu^{+} \mu^{-} \to Z' h_0 \right)$ vs. $f$, we find that for  $\sqrt{s}=4000$ GeV the largest $Z'h_0$ production is obtained, while for  $\sqrt{s}= 8000$ GeV the smallest production is reached. The following contributions are generated for these cases: $\sigma_{T}\left(\mu^{+} \mu^{-} \to Z' h_0 \right)=[3.493, 3.490]\times 10^{-2}$ fb and $\sigma_{T}\left(\mu^{+} \mu^{-} \to Z' h_0 \right)=[3.55, 3.32]\times 10^{-3}$ fb.
 On the other hand, in the  $\sigma_{T}\left(\mu^{+} \mu^{-} \to Z' h_0 \right)$ vs. $F$ scenario we find that the main $Z'h_0$ production arises again for $\sqrt{s}=4000$ GeV, $\sigma_{T}\left(\mu^{+} \mu^{-} \to Z' h_0 \right)=[4.06, 1.30] \times 10^{-2}$ fb when $F\in [3000, 4800]$ GeV. For this case, one can see from Fig.~\ref{fF-s}(b) that for values larger than $F=4800$ GeV, the mass of the new neutral gauge boson $Z'$ increases and therefore, the energy of the center-of-mass of the collider, $\sqrt{s}=4000$ GeV, is no longer sufficient for the production of $Z'h_0$.  The smallest contribution is reached for $\sqrt{s}=8000$ GeV,  $\sigma_{T}\left(\mu^{+} \mu^{-} \to Z' h_0 \right)=[4.62,2.97]\times 10^{-3}$ fb. 
 The numerical results clearly show that $\sigma_{T}\left(\mu^{+} \mu^{-} \to Z' h_0 \right)$ and $m_{Z'}$ depend mainly on the $F$ energy scale. There is little sensitivity to a change in the values of the $f$ scale.

To establish a benchmark of the $Z'h_0$ production at the future muon collider, we consider the machine design luminosities provided in Table I of  Ref.~\cite{Accettura:2023ked}: ${\cal L}_{\text{int}}=2000 \hspace{0.8mm}{\rm fb^{-1}}$ and ${\cal L}_{\text{int}}=20000 \hspace{0.8mm}{\rm fb^{-1}}$ with their respective center-of-mass energies of  $\sqrt{s}= 3000$ GeV and $\sqrt{s}= 10000$ GeV. 
In  Tables~\ref{zph13}  and~\ref{zph14}, we present an estimate of the number of expected events of the production $Z' h_0$ for different values of the new physics scales: $f=1 000$ GeV and $F=3 000$ GeV, and $f=2 000$ GeV and $F=5000$ GeV.
According to the numerical results, the BLHM's contribution to the production of $Z'h_0$ at the muon collider decreases as the mass of the new gauge boson $Z'$ increases. 
The production of the new neutral gauge boson $Z'$ and the SM Higgs boson $h_0$ seems to be promising at the future muon collider.

\begin{table}[H]
\caption{The total production of $Z' h_0$ at the future muon collider in the context of the BLHM when  $f=1 000\ \text{GeV}$ and $\ F=3000\ \text{GeV}$ ($m_{Z'} \approx 2489$ GeV).
\label{zph13}}
    \centering
    \begin{tabular}{|c|c|c|}
    \hline
     \multicolumn{3}{|c|}{ $\tan\, \beta=3$} \\
    \hline
    \multicolumn{3}{|c|}{ $f=1 000$ GeV, $F=3 000$ GeV} \\
    \hline
         $\mathcal{L}_{\text{int}}$\, [$fb^{-1}$] & $\sqrt{s}$\,  [GeV] & Expected events     \\
         \hline
         2000  & 3000 & 217   \\
         \hline
         20000 & 10000 & 50   \\
         \hline
    \end{tabular}
\end{table}

\begin{table}[H]
\caption{The total production of $Z' h_0$ at the future muon collider in the context of the BLHM when  $f=2 000\ \text{GeV}$ and $\ F=5000\ \text{GeV}$ ($m_{Z'} \approx 4240$ GeV).
\label{zph14}}
    \centering
    \begin{tabular}{|c|c|c|}
    \hline
     \multicolumn{3}{|c|}{ $\tan\, \beta=3$} \\
    \hline
    \multicolumn{3}{|c|}{ $f=2 000$ GeV, $F=5 000$ GeV} \\
    \hline
         $\mathcal{L}_{\text{int}}$\, [$fb^{-1}$] & $\sqrt{s}$\,  [GeV] & Expected events     \\
         \hline
         2000  & 3000 & 104  \\
         \hline
         20000 & 10000 & 28   \\
         \hline
    \end{tabular}
\end{table}

\subsubsection{Production of the heavy Higgs boson $H_0$ and the gauge boson $Z'$}

\begin{figure}[H] 
\center
{\includegraphics[width=10.0cm]{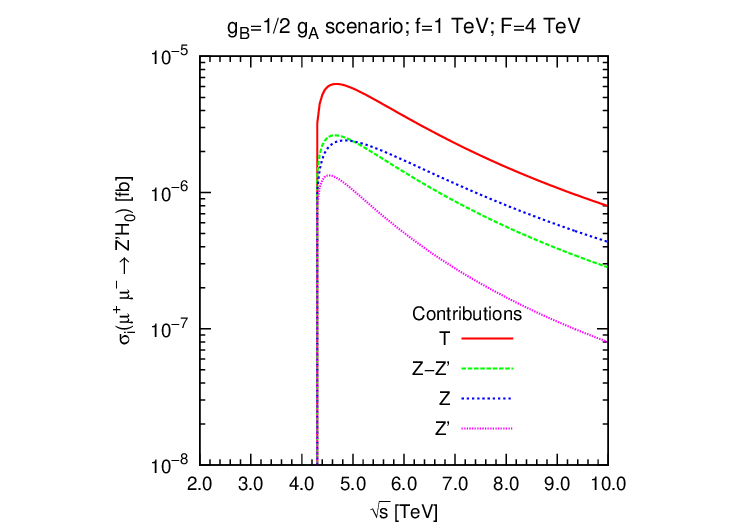}}
\caption{The different contributions to the cross section of the process  $\mu^{+}\mu^{-}\to (Z, Z') \to Z' H_0$   as a function of $\sqrt{s}$.} 
\label{S-ZH0}    
\end{figure}

We now investigate the contribution of the BLHM  to the $Z'H_0$ production cross-section, where $H_0$ represents a new heavy Higgs boson, which we have set to the order of about 1 TeV, consistently with current results from the search for new scalar bosons~\cite{ATLAS:2020gxx,CMS:2019ogx}.
The total cross section $\sigma_{T}\left(\mu^{+} \mu^{-} \to Z' H_0 \right)$ is calculated from the partial contributions provided by  $\sigma_{Z}\left(\mu^{+} \mu^{-} \to Z' H_0 \right)$, $\sigma_{Z'}\left(\mu^{+} \mu^{-} \to Z' H_0 \right)$, and $\sigma_{Z-Z'}\left(\mu^{+} \mu^{-} \to Z' H_0 \right)$ (see Eq.~(\ref{ZH0tot})). Fig.~\ref{S-ZH0} illustrates the numerical contributions of each of these when the free parameter $\sqrt{s} \in [2000,10000]$ GeV. The other input parameters are set to $\tan \beta=3$, $f=1000$ GeV, and $F=4000$ GeV. 
 As this figure shows, the $\sigma_{Z}\left(\mu^{+} \mu^{-} \to Z' H_0 \right)$ and $\sigma_{Z-Z'}\left(\mu^{+} \mu^{-} \to Z' H_0 \right)$ curves compete with each other to provide the largest contribution to $\sigma_{T}\left(\mu^{+} \mu^{-} \to Z' H_0 \right)$. In contrast, the $\sigma_{Z'}\left(\mu^{+} \mu^{-} \to Z' H_0 \right)$ curve generates a suppressed contribution throughout the study interval of the $\sqrt{s}$  parameter.
The highest point of the curves involved are reached when  $\sigma_{Z'}\left(\mu^{+} \mu^{-} \to Z' H_0 \right)=1.34 \times 10^{-6}$ fb ($\sqrt{s}\approx 4530$ GeV), $\sigma_{Z-Z'}\left(\mu^{+} \mu^{-} \to Z' H_0 \right)=2.63 \times 10^{-6}$ fb ($\sqrt{s}\approx 4650$ GeV), $\sigma_{Z}\left(\mu^{+} \mu^{-} \to Z' H_0 \right)=2.41 \times 10^{-6}$ fb ($\sqrt{s}\approx 4 840$ GeV), and $\sigma_{T}\left(\mu^{+} \mu^{-} \to Z' H_0 \right)=6.28 \times 10^{-6}$ fb ($\sqrt{s}\approx 4 670$ GeV).
The $\sigma_{i}\left(\mu^{+} \mu^{-} \to Z' H_0 \right)$ curves show a strong dependence on the $\sqrt{s}$ parameter.

\begin{figure}[H]
\center
\subfloat[]{\includegraphics[width=8.20cm]{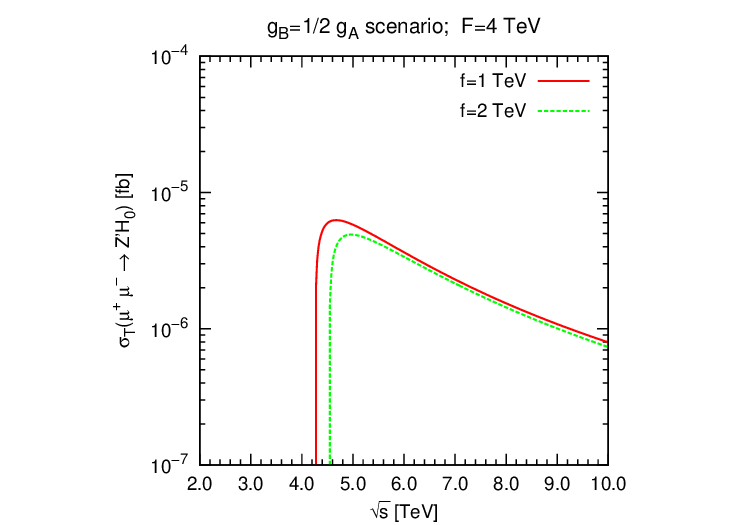}}
\subfloat[]{\includegraphics[width=8.20cm]{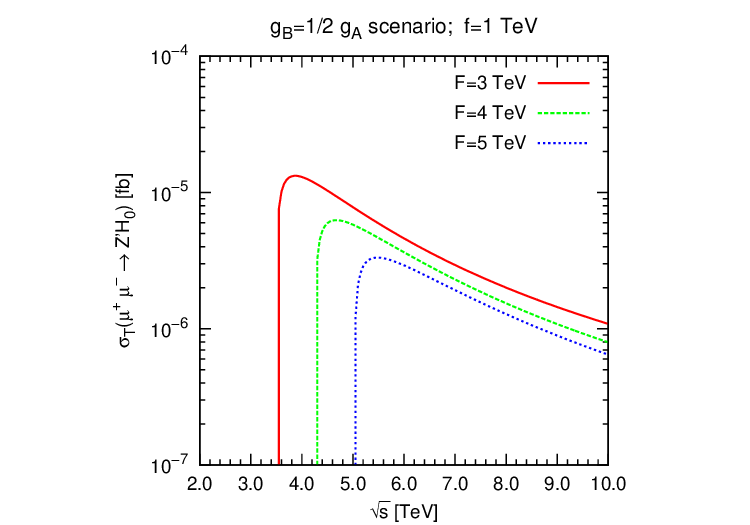}}
        \caption{The total cross section of the process $\mu^{+}\mu^{-}\to (Z, Z') \to Z' H_0$   as a function of $\sqrt{s}$.
        a)  The curves are generated for $f=1000$ GeV  and $f=2 000$ GeV.
     b)   The curves are generated for $F=3 000$ GeV,  $F=4000$ GeV, and  $F=5 000$ GeV.}
\label{SH0-12345}
\end{figure}

In Fig.~\ref{SH0-12345}, we study the impact of the parameters $\sqrt{s}$, $f$, and $F$ on the total production cross section of the process $\mu^{+} \mu^{-} \to Z'H_0$. We will first discuss the behavior observed in Fig.~\ref{SH0-12345}(a). Note that the two curves have been generated for two different values of the $f$ scale: $f=1000$ GeV and $f=2000$ GeV. The other $F$ energy scale for this plot is set to 4000 GeV. The $Z'H_0$ production cross section shows little dependence on the $f$ scale, but such dependence is more appreciable concerning the $\sqrt{s}$ parameter.
The curve providing the highest contribution is obtained when the scale $f=1000$ GeV, the highest peak is reached when $\sigma_{T}\left(\mu^{+} \mu^{-} \to Z' H_0 \right)=6.26 \times 10^{-6}$ fb for $\sqrt{s} \approx 4720$ GeV.
We now turn to discuss the dependence of $\sigma_{T}\left(\mu^{+} \mu^{-} \to Z' H_0 \right)$ on $\sqrt{s}$ and $F$ as depicted in Fig.~\ref{SH0-12345}(b).
The curves for the scales $F=3000, 4000, 5000$ GeV have been obtained. From these, we can observe that the curve generating larger numerical values arises when the $F$ scale acquires small values, particularly $F=3000$ GeV. For this case, we obtain that the highest point is reached for $\sigma_{T}\left(\mu^{+} \mu^{-} \to Z' H_0 \right)=1.33 \times 10^{-5}$ fb when $\sqrt{s} \approx 3870$ GeV. 
 Furthermore, it is evident that the value of $\sigma_{T}\left(\mu^{+} \mu^{-} \to Z' H_0 \right)$ decreases significantly as the collider center-of-mass energy, $\sqrt{s}$, increases. There is a notable correlation between $\sigma_{T}\left(\mu^{+} \mu^{-} \to Z' H_0 \right)$ and both $\sqrt{s}$ and $F$.

\begin{figure}[H] 
\center
{\includegraphics[width=10.0cm]{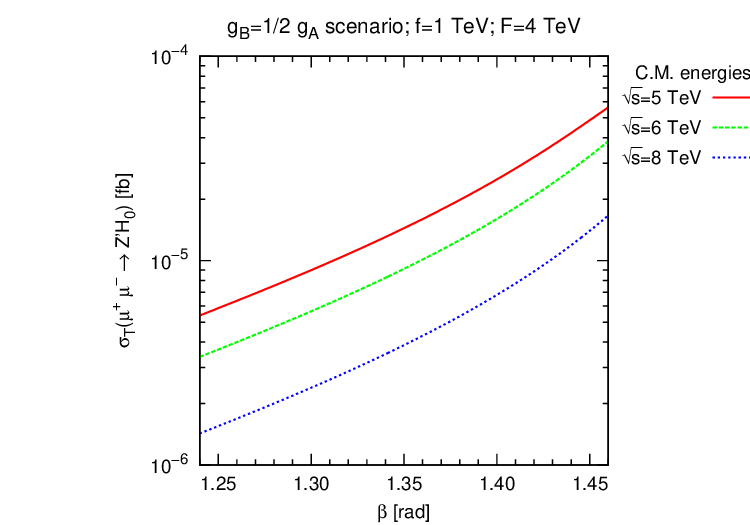}}
\caption{The total cross section of the process $\mu^{+}\mu^{-}\to (Z, Z') \to Z' H_0$   as a function of parameter $\beta$ ($\tan \beta \in [3,9]$).
 The curves are generated for different values of the center-of-mass energy $\sqrt{s}$: $\sqrt{s}=5000$ GeV, $\sqrt{s}=6000$ GeV, and $\sqrt{s}=8000$ GeV.} 
 \label{SH0vsbeta}    
\end{figure}

From the above discussions, the contributions of the BLHM to the Higgs-strahlung process depend on the free parameters $f$, $F$, $\sqrt{s}$, and $\beta$.
We therefore present the correlation between $\sigma_{T}\left(\mu^{+} \mu^{-} \to Z' H_0 \right)$ versus  $\beta$ and $\sqrt{s}$, with the scales of the new physics set to $f=1000$ GeV and $F=4000$ GeV. For this purpose, we plot $\sigma_{T}\left(\mu^{+} \mu^{-} \to Z' H_0 \right)$ as a function of the mixing angle $\beta$ for different values of the center-of-mass energy: $\sqrt{s}=5000, 6000, 8000$ GeV. 
As illustrated in Fig.~\ref{SH0vsbeta}, the curves demonstrate a notable increase of approximately one order of magnitude as $\beta$ increases up to 1.46 radians (or $\tan^{-1} (9)$).  On the other hand, $\sigma_{T}\left(\mu^{+} \mu^{-} \to Z' H_0 \right)$ obtains larger values when the parameter $\sqrt{s}=5000$ GeV:  $\sigma_{T}\left(\mu^{+} \mu^{-} \to Z' H_0 \right)=[5.81\times 10^{-6}, 5.63 \times 10^{-5}]$ fb for $\beta \in [1.25, 4.46]$ rad. Conversely, when $\sqrt{s}=8000$ GeV a small contribution is generated for the $Z'H_0$ production cross section,  $\sigma_{T}\left(\mu^{+} \mu^{-} \to Z' H_0 \right)=[1.54 \times 10^{-6}, 1.67 \times 10^{-5}]$ fb. 
Both parameters, $\beta$ and $\sqrt{s}$, have a notable impact on the  $\sigma_{T}\left(\mu^{+} \mu^{-} \to Z' H_0 \right)$ results.

\begin{figure}[H]
\center
\subfloat[]{\includegraphics[width=8.20cm]{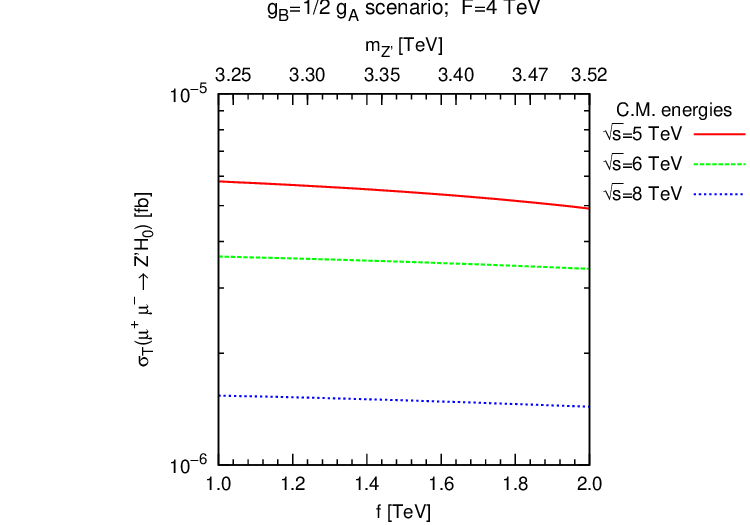}}
\subfloat[]{\includegraphics[width=8.20cm]{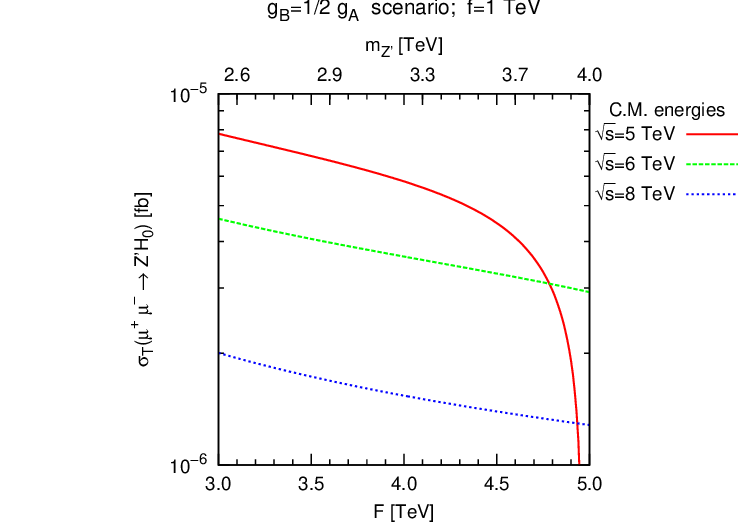}}
        \caption{a) The total cross section of the process $\mu^{+}\mu^{-}\to (Z, Z') \to Z' H_0$   as a function of the $f$ energy scale.
        b) The total cross section of the process $\mu^{+}\mu^{-}\to (Z, Z') \to Z' H_0$   as a function of the $F$ energy scale.
        The curves are generated for different values of the center-of-mass energy $\sqrt{s}$: $\sqrt{s}=5000$ GeV, $\sqrt{s}=6000$ GeV, and $\sqrt{s}=8000$ GeV.}
\label{SH0-fF}
\end{figure}

Below, we examine the dependence of $\sigma_{T}\left(\mu^{+} \mu^{-} \to Z' H_0 \right)$ on the scale of the new physics, $f$ or $F$.
 For the purposes of this analysis, $f$ lies in the interval $[1000, 2000]$ GeV and $F$ lies in the range  $[3000, 5000]$ GeV.  In the left plot of Fig.~\ref{SH0-fF}, we can observe the behavior of $\sigma_{T}\left(\mu^{+} \mu^{-} \to Z' H_0 \right)$ versus $f$ while fixed the second scale $F$ to 4000 GeV. The different curves shown have been generated for the following values of the center-of-mass energy: 5000, 6000, and 8000 GeV. In these cases, the numerical contributions of the  $Z'H_0$ production cross section are found to be $[5.81, 4.91]\times 10^{-6}$ fb, $[3.64, 3.38]\times 10^{-6}$ fb, and $[1.54, 1.43]\times 10^{-6}$ fb, respectively.
Our results show that the significant contribution is achieved when $\sqrt{s}=5000$ GeV, while the small contribution is reached when $\sqrt{s}=8000$ GeV. 
In the right plot of Fig.~\ref{SH0-fF}, we now appreciate the three curves generated for  $\sigma_{T}\left(\mu^{+} \mu^{-} \to Z' H_0 \right)$ as a function of the $F$ scale while the other scale is fixed to $f=1000$ GeV. The most significant contribution is obtained when $\sqrt{s}=5000$ GeV, which corresponds to a $Z'H_0$ production cross section of $\sigma_{T}\left(\mu^{+} \mu^{-} \to Z' H_0 \right)=[7.81, 1.15] \times 10^{-6}$ fb when $F\in[3000, 4950]$ GeV. For the region where $F>4950$ GeV, a center-of-mass energy greater than 5000 GeV would be required to produce a gauge boson $Z'$ in association with a new heavy Higgs boson $H_0$. On the other hand, the smallest contribution is reached when $\sqrt{s}=8000$ GeV, finding that $\sigma_{T}\left(\mu^{+} \mu^{-} \to Z' H_0 \right)=[2.00, 1.28] \times 10^{-6}$ fb.
Our analysis of the two scenarios, $\sigma_{T}\left(\mu^{+} \mu^{-} \to Z' H_0 \right)$ vs. $f$ and $\sigma_{T}\left(\mu^{+} \mu^{-} \to Z' H_0 \right)$ vs. $F$, indicates that $\sigma_{T}\left(\mu^{+} \mu^{-} \to Z' H_0 \right)$ is dependent on the energy scales $f$ and $F$, with a greater impact observed on the $F$ scale. This same sensitivity is evident in the parameter $m_{Z'}$.

\begin{figure}[H]  
\center
{\includegraphics[width=10.0cm]{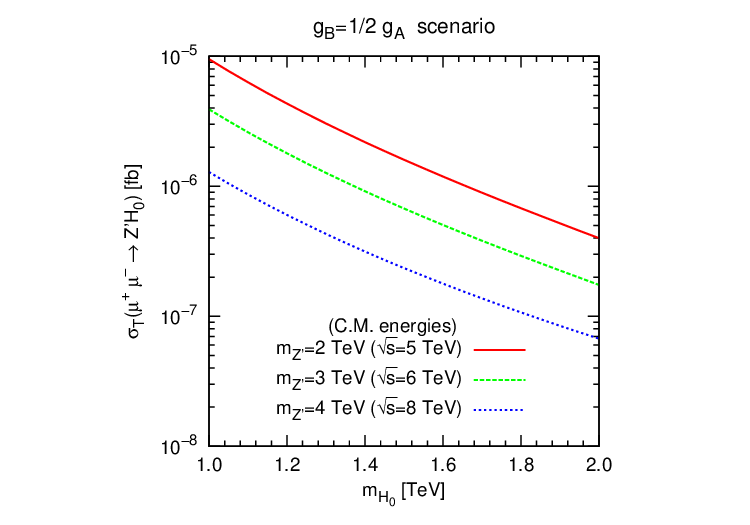}}
\caption{The total cross section of the process $\mu^{+}\mu^{-}\to (Z, Z') \to Z' H_0$  as a function of the Higgs boson mass $H_0$. 
 The curves are generated for different values of the mass of the gauge boson $Z'$.}  
 \label{SH0mass}   
\end{figure}

Fig.~\ref{SH0mass} illustrates the behavior of the $Z'H_0$ production cross section as a function of the mass of the new heavy Higgs boson $H_0$. We have chosen three different values for the mass of the gauge boson $Z'$.
For the generated curves we find that their numerical contributions to the $\sigma_{T}\left(\mu^{+} \mu^{-} \to Z' H_0 \right)$ cross section are $\sigma_{T}\left(\mu^{+} \mu^{-} \to Z' H_0 \right)=[9.50 \times 10^{-6}, 3.99 \times 10^{-7}]$ fb ($m_{Z'}=2000$ GeV), $\sigma_{T}\left(\mu^{+} \mu^{-} \to Z' H_0 \right)=[3.91 \times 10^{-6}, 1.74 \times 10^{-7}]$ fb ($m_{Z'}= 3000$ GeV), and $\sigma_{T}\left(\mu^{+} \mu^{-} \to Z' H_0 \right)=[1.29 \times 10^{-6}, 6.73 \times 10^{-8}]$ fb ($m_{Z'}= 4000$ GeV) when $m_{H_0} \in [1000,2000]$ GeV. 
The $Z'H_0$ production cross section decreases by one or two orders of magnitude as $m_{H_0}$ increases up to 2000 GeV. It is also evident that a higher center-of-mass energy is required to produce the gauge boson $Z'$ and the new Higgs boson $H_0$ as their masses increase.
We have numerically calculated the $\sigma_{T}\left(\mu^{+} \mu^{-} \to Z' H_0 \right)$ cross section as a function of the $m_{H_0}$ parameter. From Fig.~\ref{SH0mass}, it is clear that the region with the greater predictive importance corresponds to $m_{H_0}$ around 1000 GeV, in particular, when $m_{Z'}=2000$ GeV. The $\sigma_{T}\left(\mu^{+} \mu^{-} \to Z' H_0 \right)$  cross section also shows a strong dependence on the $m_{H_0}$ and $m_{Z'}$ parameters.

As performed in Subsection~\ref{production-zp-h0}, it is possible to estimate the expected number of events of the $Z' H_0$ production at a future muon collider. For this purpose, we consider the muon collider's expected integrated luminosity in the operation's first or second stage. In this experimental scenario, observing any event related to the $\mu^{+} \mu^{-} \to Z' H_0 $ process would be challenging. This process provides a suppressed cross-section, increasing the difficulty of observation.

\subsection{Production of the Higgs bosons $h_0$, $H_0$ and the gauge boson $Z'$ under the $g_{B}=g_{A}$ scenario} \label{scenarioBequalA}

In this Subsection, we examine the production of the Higgs boson $h_0$ or $H_0$ and the new heavy boson $Z'$ in the scenario where the gauge couplings $g_B$ and $g_{A}$ satisfy the condition $g_B=g_A$. In the same scenario, we also study the partial decay widths and branching ratios of the neutral gauge boson $Z'$.  

\subsubsection{$\Gamma_{Z^{\prime}} $ }

\begin{figure}[H]
\center
\subfloat[]{\includegraphics[width=7.0cm]{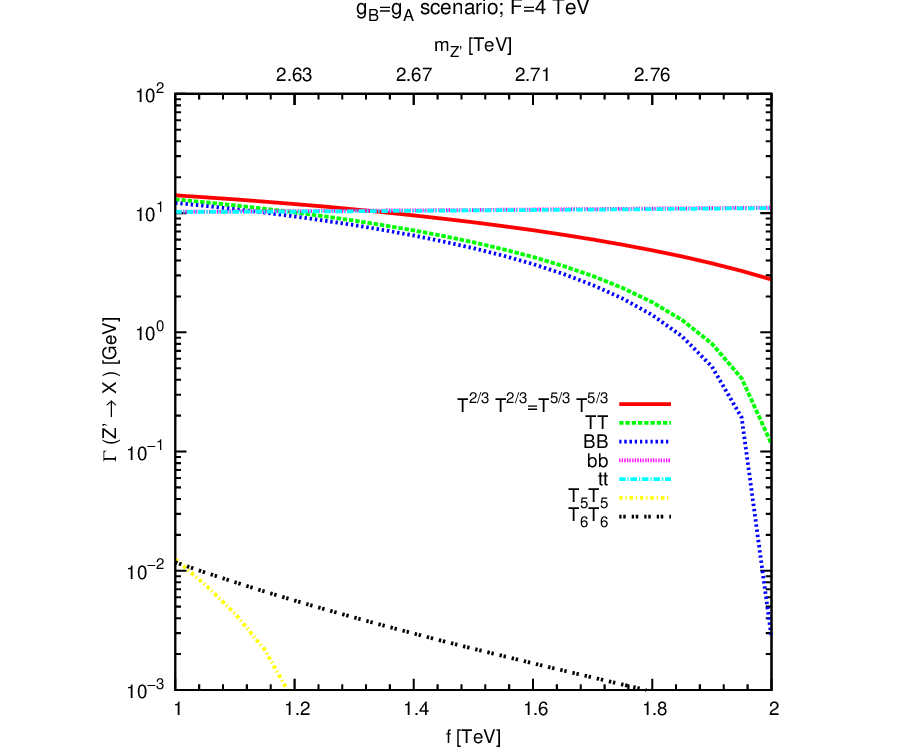}}
\subfloat[]{\includegraphics[width=7.0cm]{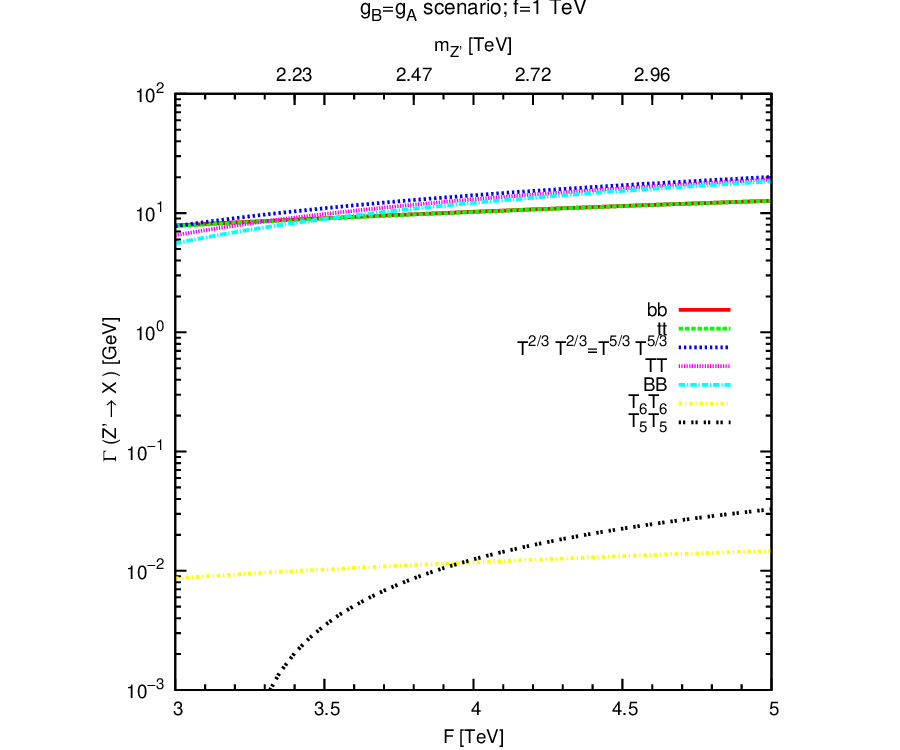}}
\caption{ Decay widths for the processes  $Z' \to X$  where $X=b\bar b, B\bar B, t\bar t,  T\bar T, T_5\bar T_5,$ $ T_6\bar T_6, T^{2/3}\bar T^{2/3}, T^{5/3}\bar T^{5/3}$. a) $\Gamma(Z^{\prime} \to X)$ as a function of the $f$ energy scale (with  $F=4 000$ GeV). b) $\Gamma(Z^{\prime}  \to X)$ as a function of the $F$ energy scale (with $f=1 000$ GeV). }
\label{widhtsAB}
\end{figure}

In the framework of the BLHM, where the gauge couplings are equal, the gauge boson $Z'$ is observed to decay primarily to fermions: $b\bar b, B\bar B, t\bar t,  T\bar T, T_5\bar T_5,$ $ T_6\bar T_6, T^{2/3}\bar T^{2/3}, T^{5/3}\bar T^{5/3}$ (see Fig.~\ref{widhtsAB}). For these processes, we have calculated the decay widths, denoted as $\Gamma(Z'\to X)$ ($X \equiv b\bar b, B\bar B, t\bar t,  T\bar T, T_5\bar T_5,$ $ T_6\bar T_6, T^{2/3}\bar T^{2/3}, T^{5/3}\bar T^{5/3}$). Fig.~\ref{widhtsAB} depicts the variation of the decay width as a function of the energy scale, $f$ or $F$. The respective analysis intervals for the energy scales are $f\in[1000,2000]$ GeV and $F\in[3000,5000]$ GeV.  
The generated curves in the left plot of  Fig.~\ref{widhtsAB} have been obtained by setting $\tan \beta=3$ and $F = 4000$ GeV. For this case, we can observe that the main contribution is given by the $Z'\to T^{2/3}\bar T^{2/3}$ and $Z'\to T^{5/3}\bar T^{5/3}$  decays while $f\in [1000,1320]$ GeV:
$\Gamma(Z'\to T^{2/3}\bar T^{2/3})=\Gamma(Z'\to T^{5/3}\bar T^{5/3})=[14.11, 10.39]$ GeV.
For the region $f\in (1320,2000]$ GeV, the $Z'\to b\bar b$ process generates the largest numerical contribution, $\Gamma(Z'\to  b\bar b) =(10.39,11.12]$ GeV.
 In contrast, the smallest contribution is provided by the $Z'\to T_5 \bar T_5$ process whose numerical contribution in the interval $f= [1000, 1200]$ GeV is $\Gamma(Z'\to T_5\bar T_5) = [1.25 \times 10^{-2}, 1.00 \times 10^{-3}]$ GeV.
On the other hand, in the right plot of Fig.~\ref{widhtsAB}, we illustrate the behavior of $\Gamma(Z'\to X)$ as a function of the $F$ scale while maintaining the other $f$ scale at 1000 GeV. In this scenario, we can see that the curves providing the most significant contributions in the $f$ scale study range correspond to the $Z'\to b\bar b$, $Z'\to T^{2/3}\bar T^{2/3}$, and  $Z'\to T^{5/3}\bar T^{5/3}$ decays: $\Gamma(Z'\to  b\bar b) =[7.86,7.93]$ GeV while $f\in [3000,3030]$ GeV, and   $\Gamma(Z'\to T^{2/3}\bar T^{2/3})=\Gamma(Z'\to T^{5/3}\bar T^{5/3})=(7.93,20.08]$ GeV when  $f\in (3030,5000]$ GeV.
  In contrast, the smallest curve is given by the $Z'\to T_5\bar T_5$ and $Z'\to T_6\bar T_6$ processes:  $\Gamma(Z'\to  T_5\bar T_5) = [1.05\times 10^{-3}, 1.15 \times 10^{-2}]$ GeV for $F= [3320,3950]$ GeV, and $\Gamma(Z'\to  T_6\bar T_6) = ( 1.15, 1.47] \times 10^{-2}$ GeV for $F= (3950,5000]$ GeV.
 As can be seen on the upper horizontal axis of these plots, the gauge boson $Z'$ mass varies as the scales of new physics are altered within the established range of analysis. Notably, the  $m_{Z'}$  parameter demonstrates a stronger correlation with the $F$ scale than the $f$ scale. 
 The partial decay widths $\Gamma(Z'\to X)$  also show a sensitivity to changes in the new physics scales, $f$ and $F$.

\subsubsection{$\text{Br}\left(Z^{\prime} \to X \right)$}

\begin{figure}[H]
\center
\subfloat[]{\includegraphics[width=8.0cm]{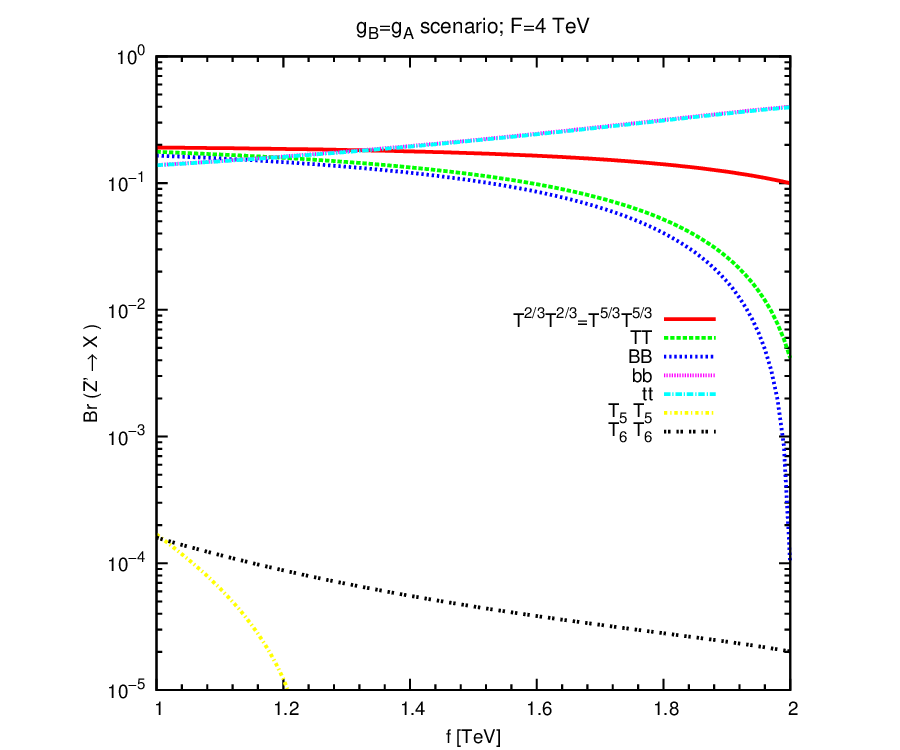}}
\subfloat[]{\includegraphics[width=8.0cm]{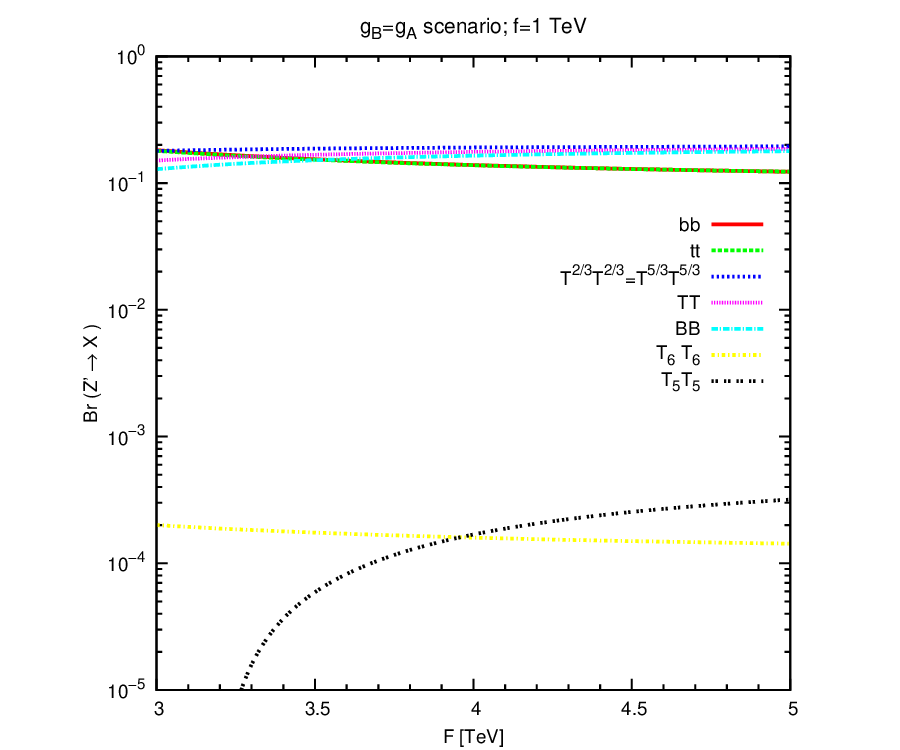}}
\caption{ \label{branchingAB} Branching ratios for the processes $Z' \to X$  where $X= b\bar b, B\bar B, t\bar t,  T\bar T, T_5\bar T_5,$ $ T_6\bar T_6, T^{2/3}\bar T^{2/3}, T^{5/3}\bar T^{5/3}$. a) $\text{Br}(Z^{\prime} \to X)$ as a function of the $f$ energy scale (with  $F=4 000$ GeV). b) $\text{Br}(Z^{\prime}  \to X)$ as a function of the $F$ energy scale (with  $f=1 000$ GeV).
}
\end{figure}

 We have calculated the corresponding $\text{Br}(Z^{\prime}  \to X)$  branching ratios for the fermionic decays of the gauge boson $Z'$. These are illustrated in Fig.~\ref{branchingAB}. Fig.~\ref{branchingAB}(a) shows the behaviour of $\text{Br}(Z^{\prime}  \to X)$ versus the $f$ scale. From this figure we appreciate that the largest contributions to $\text{Br}(Z^{\prime}  \to X)$ are given by the $Z' \to  T^{2/3}\bar T^{2/3}$, $Z' \to   T^{5/3}\bar T^{5/3}$, and $Z' \to  b\bar b$  decays: 
$\text{Br}(Z^{\prime}  \to  T^{2/3}\bar T^{2/3})=\text{Br}(Z^{\prime}  \to   T^{5/3}\bar T^{5/3})=[1.91,1.82] \times 10^{-1}$ when $f\in [1000,1320]$ GeV, and 
$\text{Br}(Z^{\prime}  \to  b\bar b)=(1.82,4.00] \times 10^{-1}$ when $f\in (1320,2000]$ GeV. Our numerical analysis shows us that the $Z'\to T_5\bar T_5$ process generates the smallest contribution to $\text{Br}(Z^{\prime}  \to X)$: $\text{Br}(Z^{\prime}  \to   T_5\bar T_5)=[1.69 \times 10^{-4}, 1.22 \times 10^{-5}]$ for $f\approx [1000, 1200]$ GeV.
Complementarily, we also study the behavior of $\text{Br}(Z^{\prime}  \to X)$ versus the $F$ scale. For this case, the largest contributions arise through the  $Z'\to b\bar b$, $Z'\to T^{2/3}\bar T^{2/3}$, and  $Z'\to T^{5/3}\bar T^{5/3}$ decays:
$\text{Br}(Z^{\prime}  \to  b\bar b)=[1.82, 1.79] \times 10^{-1}$ when $F\in [3000,3030]$ GeV, and 
$\text{Br}(Z^{\prime}  \to  T^{2/3}\bar T^{2/3})=\text{Br}(Z^{\prime}  \to   T^{5/3}\bar T^{5/3})=(1.79,1.95] \times 10^{-1}$ when $F\in (3030,5000]$ GeV.
The suppressed contribution is reached when $\text{Br}(Z^{\prime}  \to  T_5\bar T_5)=[1.22 \times 10^{-5}, 1.59 \times 10^{-4}]$ while $F = [3280,3950]$ GeV, outside this interval, when $F= (3950, 5000]$ GeV, the small contribution is obtained for $\text{Br}(Z^{\prime}  \to  T_6\bar T_6)=( 1.59, 1.43] \times 10^{-4}$.
The numerical data shows that the branching ratio strongly depends on the energy scales ($f$, $F$).

 \subsubsection{ Production of the Higgs boson $h_0$ and the new gauge boson $Z'$}
 
\begin{figure}[H]
\center
\subfloat[]{\includegraphics[width=8.0cm]{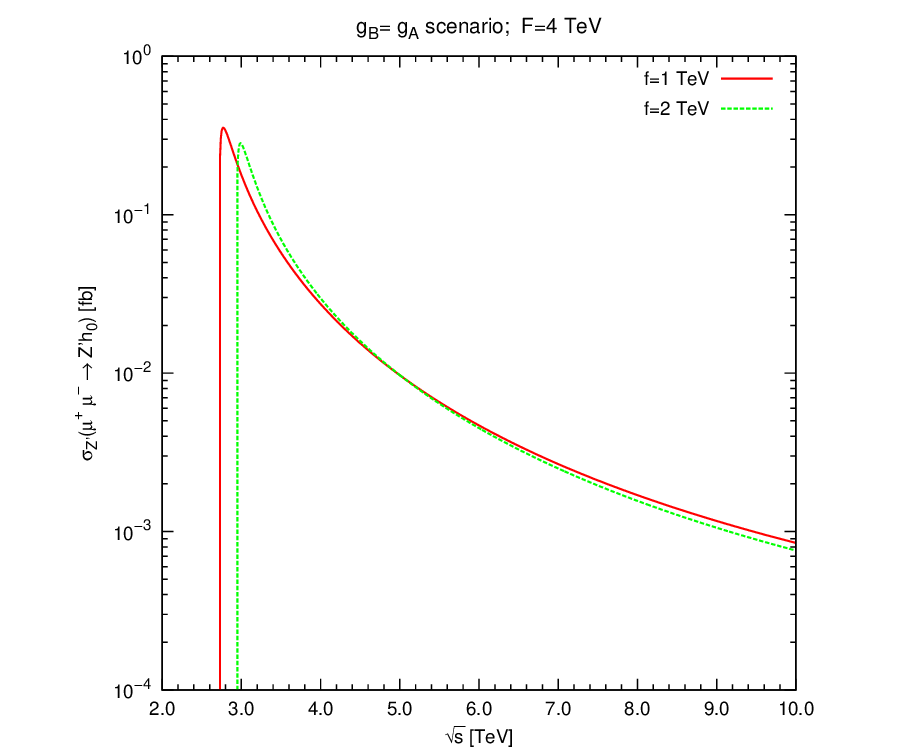}}
\subfloat[]{\includegraphics[width=8.0cm]{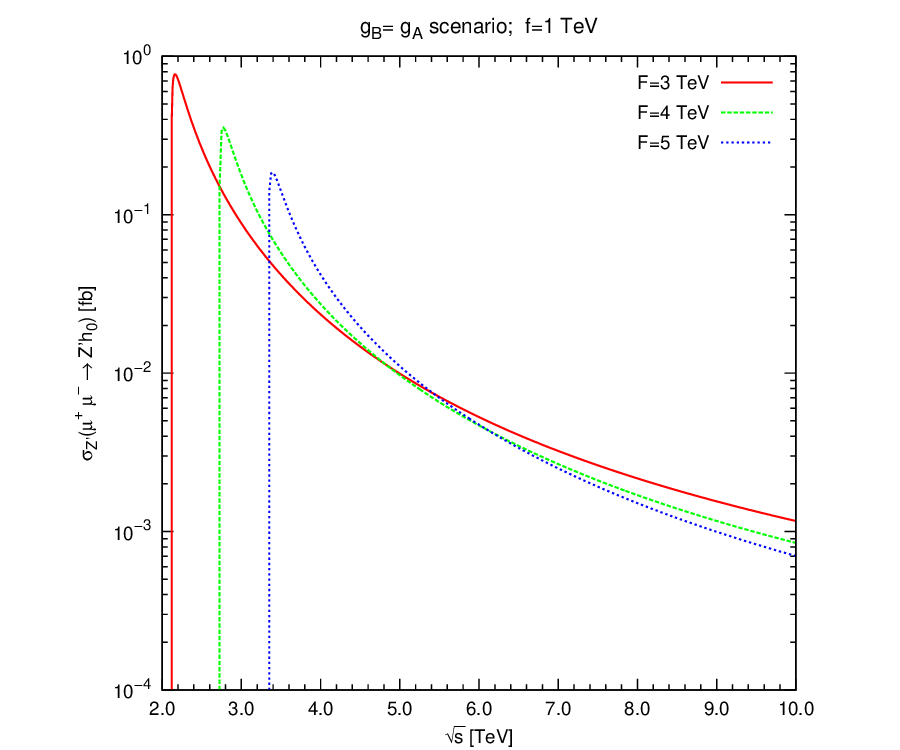}}
        \caption{Cross section of the process $\mu^{+}\mu^{-}\to  Z' \to Z' h_0$   as a function of $\sqrt{s}$.
        a)  The curves are generated for $f=1000$ GeV and $f=2 000$ GeV.
     b)   The curves are generated for $F=3 000$ GeV,  $F=4000$ GeV, and  $F=5 000$ GeV.}
\label{Psh-12-1345}
\end{figure}

 The total cross section of the Higgs-strahlung process $\mu^{+}\mu^{-}\to (Z, Z') \to Z' h_0$ is calculated using the  $\sigma_{Z}\left(\mu^{+} \mu^{-} \to Z' h_0 \right)$,  $\sigma_{Z'}\left(\mu^{+} \mu^{-} \to Z' h_0 \right)$, and  $\sigma_{Z-Z'}\left(\mu^{+} \mu^{-} \to Z' h_0 \right)$ contributions (see Eq.~(\ref{XSTotEq})).  However, in the scenario of our study where the gauge couplings $g_B$ and $g_A$ are equal, the total cross section $\sigma_{T}\left(\mu^{+} \mu^{-} \to Z' h_0 \right)$ is determined solely by the contribution from the new gauge boson $Z'$. This is due to the fact that $g_B=g_A$, which leads to $\sin \theta_g=\cos \theta_g$ ($s_g=c_g$). This condition nullifies specific Feynman rules for interaction vertices involved in our calculations, namely the $ZZ'h_0$ coupling (see Table~\ref{scalar-gauge1}). 
 In this way, in Fig.~\ref{Psh-12-1345} we show the behavior of $\sigma_{Z'}\left(\mu^{+} \mu^{-} \to Z' h_0 \right)$ as a function of the center-of-mass energy $\sqrt{s}$ for different values of the $f$ or $F$ energy scale.
  The two curves shown in Fig.~\ref{Psh-12-1345}(a) have been generated for $f=1000$ GeV and $f=2000$ GeV, while  $F=4000$ GeV and $\tan \beta=3$. Our analysis indicates that the slightly more significant contribution is generated for $f=1000$ GeV, reaching the highest point at $\sqrt{s}\approx 2770$ GeV with $\sigma_{Z'}\left(\mu^{+} \mu^{-} \to Z' h_0 \right)=3.54\times 10^{-1}$ fb.
  We will now analyze the curves generated in  Fig.~\ref{Psh-12-1345}(b), which show the behavior of $\sigma_{Z'}\left(\mu^{+} \mu^{-} \to Z' h_0 \right)$ versus $\sqrt{s}$  for different values of the $F$ energy scale while keeping fixed the scale $f=1000$ GeV.  As can be seen in this figure, the curve that reaches the most emergent production cross section occurs when $F=3000$ GeV, obtaining  $\sigma_{Z'}\left(\mu^{+} \mu^{-} \to Z' h_0 \right)=7.71\times 10^{-1}$ fb for $\sqrt{s}=2160$ GeV.
  The  $Z'h_0$ production cross-section is sensitive to parameter changes $\sqrt{s}$, $f$, and $F$. As $\sqrt{s}$ increases, the cross-section $\sigma_{Z'}\left(\mu^{+} \mu^{-} \to Z' h_0 \right)$ decreases. The same effect is present for large values of the new physics scales.

\begin{figure}[H]  
\center
{\includegraphics[width=8.0cm]{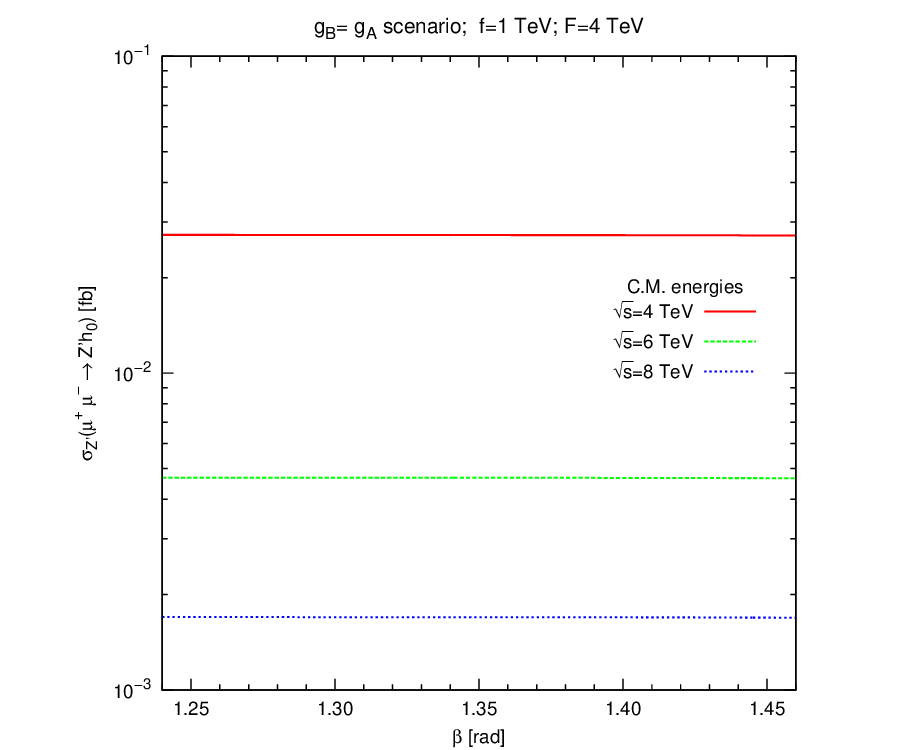}}
 \caption{Cross section of the process $\mu^{+}\mu^{-}\to  Z' \to Z' h_0$   as a function of $\beta$ parameter  ($\tan \beta \in [3,9]$).
 The curves are generated for different values of the center-of-mass energy $\sqrt{s}$: $\sqrt{s}=4000$ GeV, $\sqrt{s}=6000$ GeV, and $\sqrt{s}=8000$ GeV.}
 \label{Pshbeta-apendixx}
\end{figure}

Fig.~\ref{Pshbeta-apendixx} illustrates that the variations weakly influence the $Z'h_0$ production cross section in the $\beta$ parameter. This plot depicts the impact of varying the $\beta$ parameter from $1.24 $ to $1.46$ rad for three center-of-mass energies, i.e. $\sqrt{s} = 4000$ GeV, $\sqrt{s} = 6000$ GeV,  and $\sqrt{s} = 8000$ GeV. In this scenario, the numerical contributions obtained are   $\sigma_{Z'}\left(\mu^{+} \mu^{-} \to Z' h_0 \right)=[2.73,2.71]\times 10^{-2}$ fb ($\sqrt{s} = 4000$ GeV),   $\sigma_{Z'}\left(\mu^{+} \mu^{-} \to Z' h_0 \right)=[4.68, 4.65]\times 10^{-3}$ fb ($\sqrt{s} = 6000$ GeV), and   $\sigma_{Z'}\left(\mu^{+} \mu^{-} \to Z' h_0 \right)=[1.70,1.69]$ fb ($\sqrt{s} = 8000$ GeV). 
  While Fig.~\ref{Pshbeta-apendixx} has been generated for the values of $f=1000$ GeV and $F=4000$ GeV, the depicted behavior of  $\sigma_{Z'}\left(\mu^{+} \mu^{-} \to Z' h_0 \right)$ is analogous for any selection of $f$ and $F$, provided that it falls within the permitted intervals.

\begin{figure}[H]
\center
\subfloat[]{\includegraphics[width=8.0cm]{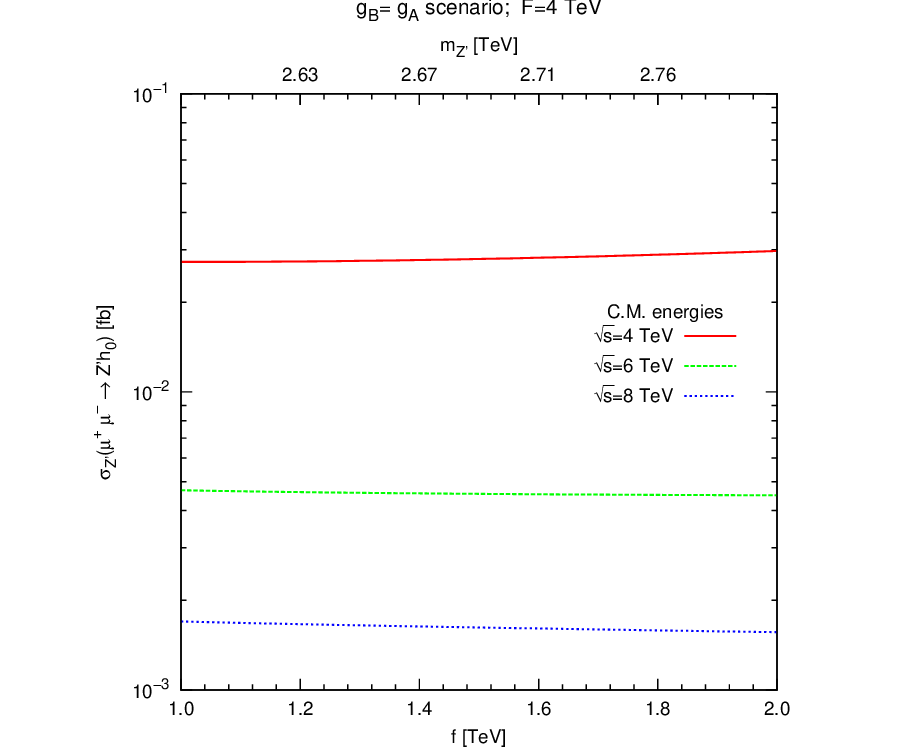}}
\subfloat[]{\includegraphics[width=8.0cm]{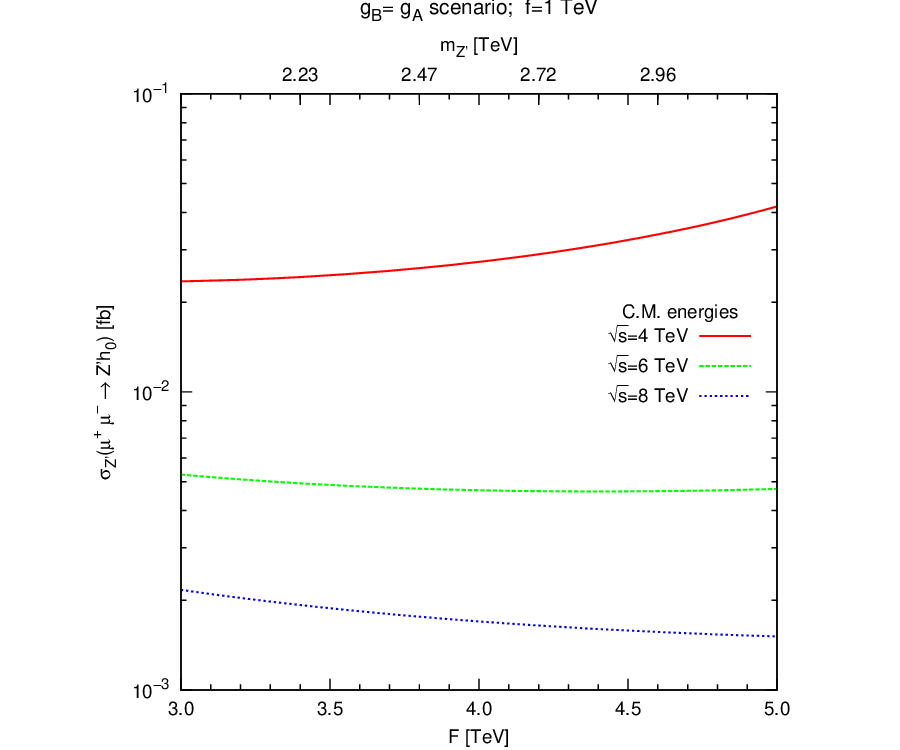}}
        \caption{ a) Cross section of the process $\mu^{+}\mu^{-}\to  Z' \to Z' h_0$   as a function of the $f$ energy scale.
        b) Cross section of the process $\mu^{+}\mu^{-}\to Z' \to Z' h_0$   as a function of the $F$ energy scale.
        The curves are generated for different values of the center-of-mass energy $\sqrt{s}$: $\sqrt{s}=4000$ GeV, $\sqrt{s}=6000$ GeV, and $\sqrt{s}=8000$ GeV.}
\label{Psh-fFF-apendixx}
\end{figure}

We now present the analysis of the generated contributions of $\sigma_{Z'}\left(\mu^{+} \mu^{-} \to Z' h_0 \right)$ when it is a function of the scale of the new physics $f$ or $F$. Fig.~\ref{Psh-fFF-apendixx}(a) shows the three curves generated for different values of the $\sqrt{s}$ parameter: $\sqrt{s}=4000, 6000, 8000$ GeV. The numerical contributions obtained for $\sigma_{Z'}\left(\mu^{+} \mu^{-} \to Z' h_0 \right)$ are 
$\sigma_{Z'}\left(\mu^{+} \mu^{-} \to Z' h_0 \right)=[2.73,2.97]\times 10^{-2}$ fb (for $\sqrt{s}=4000$ GeV),
 $\sigma_{Z'}\left(\mu^{+} \mu^{-} \to Z' h_0 \right)=[4.68, 4.50]\times 10^{-3}$ fb (for $\sqrt{s}=6000$ GeV),
  and $\sigma_{Z'}\left(\mu^{+} \mu^{-} \to Z' h_0 \right)=[1.70, 1.56] \times 10^{-3}$ fb (for $\sqrt{s}=8000$ GeV) when $f \in [1000,2000]$ GeV. 
 Fig.~\ref{Psh-fFF-apendixx}(b) illustrates the behaviour of $\sigma_{Z'}\left(\mu^{+} \mu^{-} \to Z' h_0 \right)$ as a function of the $F$ scale. The curves shown have been obtained using the same values of the $\sqrt{s}$ parameter as previously mentioned. In this instance, the  production cross-section of $Z' h_0$ is as follows:
  $\sigma_{Z'}\left(\mu^{+} \mu^{-} \to Z' h_0 \right)=[2.35, 4.19]\times 10^{-2}$ fb (for $\sqrt{s}=4000$ GeV),
   $\sigma_{Z'}\left(\mu^{+} \mu^{-} \to Z' h_0 \right)=[5.28,4.73]\times 10^{-3}$ fb (for $\sqrt{s}=6000$ GeV),
    and $\sigma_{Z'}\left(\mu^{+} \mu^{-} \to Z' h_0 \right)=[2.17, 1.51]\times 10^{-3}$ fb (for $\sqrt{s}=8000$ GeV) for $F\in [3000, 5000]$ GeV.
 It is clear that for the two cases under discussion, $\sigma_{Z'}\left(\mu^{+} \mu^{-} \to Z' h_0 \right)$ obtains larger numerical values for smaller center-of-mass energies, particularly for $\sqrt{s}=4000$ GeV. $\sigma_{Z'}\left(\mu^{+} \mu^{-} \to Z' h_0 \right)$ also shows a dependence on the energy scales $f$ and $F$, with the dependence being stronger on the $F$ scale. The same effect can be seen in the values of the mass of the heavy gauge boson $Z'$.

As implemented in Subsection~\ref{production-zp-h0}, we use the expected integrated luminosities and center-of-mass energies of the future muon collider to determine the expected number of events of the $Z' h_0$ production at that collider. For this purpose, we provide an estimate of the number of $Z'h_0$ events for certain values of the new physics scales in Tables~\ref{zph13-apendixx} and~\ref{zph14-apendixx}. These values are $f=1000$ GeV and $F=3000$ GeV, and $f=2000$ GeV and $F=5000$ GeV.

\begin{table}[H]
\caption{The total production of $Z' h_0$ at the future muon collider in the context of the BLHM when  $f=1 000\ \text{GeV}$ and $\ F=3000\ \text{GeV}$ ($m_{Z'} \approx 1990$ GeV).
\label{zph13-apendixx}}
    \centering
    \begin{tabular}{|c|c|c|}
    \hline
     \multicolumn{3}{|c|}{ $\tan\, \beta=3$} \\
    \hline
    \multicolumn{3}{|c|}{ $f=1 000$ GeV, $F=3 000$ GeV} \\
    \hline
         $\mathcal{L}_{\text{int}}$\, [$fb^{-1}$] & $\sqrt{s}$\,  [GeV] & Expected events     \\
         \hline
         2000  & 3000 & 176   \\
         \hline
         20000 & 10000 & 23   \\
         \hline
    \end{tabular}
\end{table}

\begin{table}[H]
\caption{The total production of $Z' h_0$ at the future muon collider in the context of the BLHM when  $f=2 000\ \text{GeV}$ and $\ F=5000\ \text{GeV}$ ($m_{Z'} \approx 3392 $ GeV).
\label{zph14-apendixx}}
    \centering
    \begin{tabular}{|c|c|c|}
    \hline
     \multicolumn{3}{|c|}{ $\tan\, \beta=3$} \\
    \hline
    \multicolumn{3}{|c|}{ $f=2 000$ GeV, $F=5 000$ GeV} \\
    \hline
         $\mathcal{L}_{\text{int}}$\, [$fb^{-1}$] & $\sqrt{s}$\,  [GeV] & Expected events     \\
         \hline
         2000  & 3000 & 310  \\
         \hline
         20000 & 10000 & 13   \\
         \hline
    \end{tabular}
\end{table}

\subsubsection{Production of the heavy Higgs boson $H_0$ and the new gauge boson $Z'$}

\begin{figure}[H]
\center
\subfloat[]{\includegraphics[width=8.0cm]{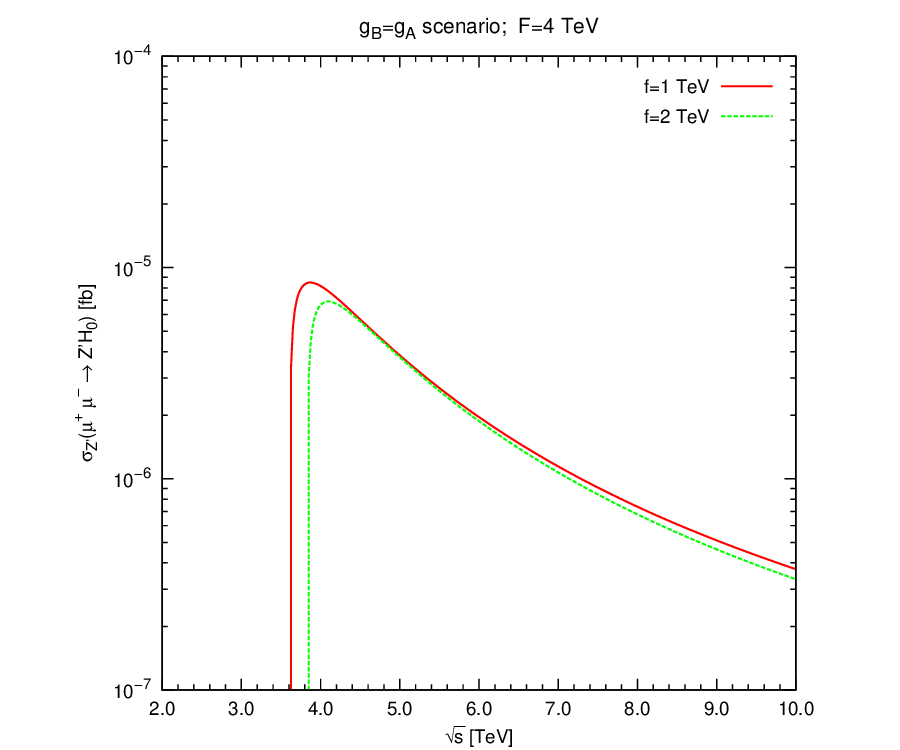}}
\subfloat[]{\includegraphics[width=8.0cm]{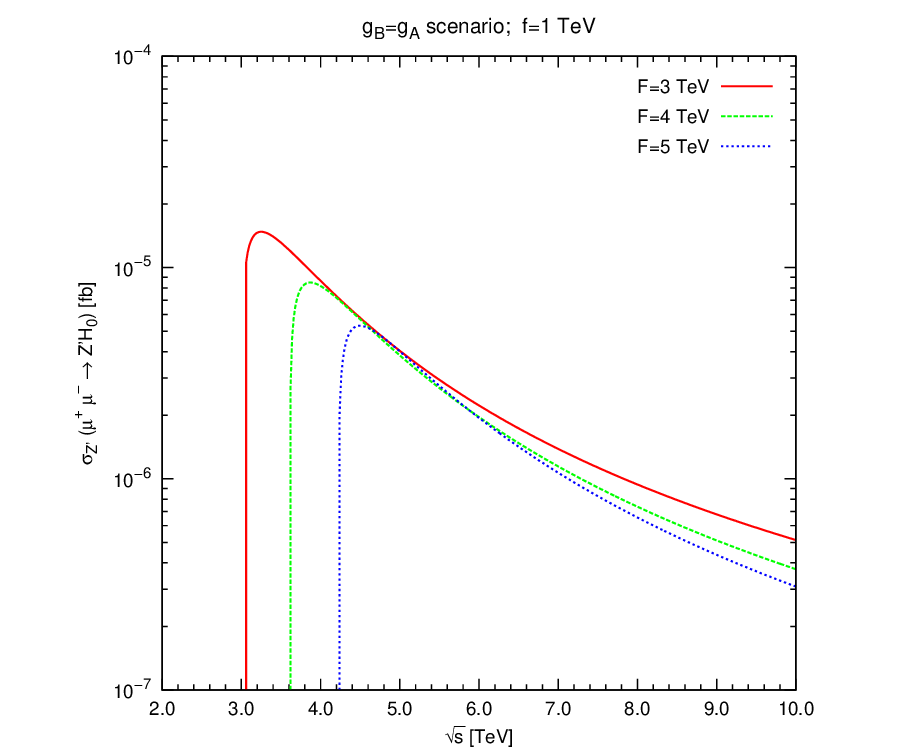}}
\caption{Cross section of the process $\mu^{+}\mu^{-}\to Z' \to Z' H_0$  as a function of $\sqrt{s}$.  
   a)  The curves are generated for $f=1000$ GeV and $f=2 000$ GeV.
   b)   The curves are generated for $F=3 000$ GeV,  $F=4000$ GeV, and  $F=5 000$ GeV.}
\label{PSH012-1345-ap}
\end{figure}

The Higgs-strahlung production process $\mu^{+}\mu^{-}\to (Z, Z') \to Z' H_0$ also reduces to the calculation of the  $\mu^{+}\mu^{-}\to Z' \to Z' H_0$ process in the scenario where the gauge couplings satisfy the $g_B=g_A$ condition. Thus, the production cross-section of $Z'H_0$ will only receive the contribution of the new gauge boson $Z'$, which will be our object of study in this Subsection. 
In Fig.~\ref{PSH012-1345-ap}, we analyze the dependence of the $ \sigma_{Z'}(\mu^{+}\mu^{-} \to Z' H_0)$ cross-section on variations of the center-of-mass energy, which takes values from 2000 to 10000 GeV. 
The curves shown in Fig.~\ref{PSH012-1345-ap}(a) have been obtained for two different values of the $f$ scale, $f=1000$ GeV and $f=2000$ GeV. In both cases, the other scale has been set at 4000 GeV.  In this figure we can see that the slightly larger numerical contributions for $ \sigma_{Z'}(\mu^{+}\mu^{-} \to Z' H_0)$ are obtained for the smaller value of the $f$ scale, i.e. for $f=1000$ GeV. The numerical contribution of $ \sigma_{Z'}(\mu^{+}\mu^{-} \to Z' H_0)$ obtained in the interval $\sqrt{s} \approx  [3 620,10 000]$ GeV is $ \sigma_{Z'}(\mu^{+}\mu^{-} \to Z' H_0)=[2.68 \times 10^{-6}, 3.73 \times 10^{-7}]$ fb.
Concerning Fig.~\ref{PSH012-1345-ap}(b), the curves have been generated for different choices of the $F$ scale values, while the $f$ scale is set to 1000 GeV in all cases.  The largest contributions for $ \sigma_{Z'}(\mu^{+}\mu^{-} \to Z' H_0)$ are obtained when $F=3000$ GeV: $ \sigma_{Z'}(\mu^{+}\mu^{-} \to Z' H_0)= [3.46 \times 10^{-6}, 5.13 \times 10^{-7}] $ fb when $\sqrt{s} \approx  [3010,10 000]$ GeV.
From the numerical data, it is evident that $ \sigma_{Z'}(\mu^{+}\mu^{-} \to Z' H_0)$ shows a dependence on the input parameters $f$, $F$, and $\sqrt{s}$.

\begin{figure}[H] 
\center
{\includegraphics[width=8.0cm]{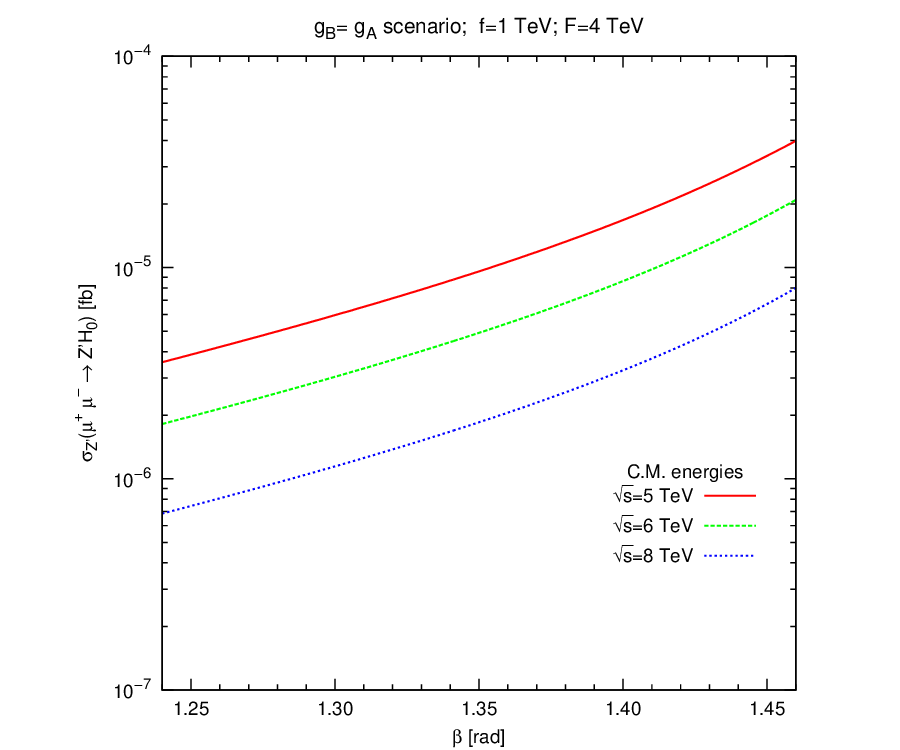}}
 \caption{Cross section of the process $\mu^{+}\mu^{-}\to  Z' \to Z' H_0$   as a function of $\beta$ parameter ($\tan \beta \in [3,9]$).
 The curves are generated for different values of the center-of-mass energy $\sqrt{s}$: $\sqrt{s}=5000$ GeV, $\sqrt{s}=6000$ GeV, and $\sqrt{s}=8000$ GeV.}
\label{PSH0beta-ap}
\end{figure}

We now analyze the sensitivity of the $Z'H_0$ production cross section concerning the $\beta$ angle and the $\sqrt{s}$ parameter while fixing the new physics scales to certain specific values.  In this way, in Fig.~\ref{PSH0beta-ap}, we show the curves generated for three different values of the $\sqrt{s}$ parameter: $\sqrt{s}=5000, 6000, 8000$ GeV.
 From the figure we can also see that the main contribution for $ \sigma_{Z'}(\mu^{+}\mu^{-} \to Z' H_0)$ is obtained for $\sqrt{s}=5000$ GeV with $ \sigma_{Z'}(\mu^{+}\mu^{-} \to Z' H_0)=[3.84\times 10^{-6}, 3.98 \times 10^{-5}] $ fb when $\beta  \in [1.25, 1.46]$ radians.
 All the curves involved increase about one order of magnitude as $\beta$ increases to 1.46 radians.
As shown in the corresponding figure, the $ \sigma_{Z'}(\mu^{+}\mu^{-} \to Z' H_0)$ cross section demonstrates a notable sensitivity to variations in the $\beta$ and $\sqrt{s}$ parameters.

\begin{figure}[H]
\center
\subfloat[]{\includegraphics[width=8.0cm]{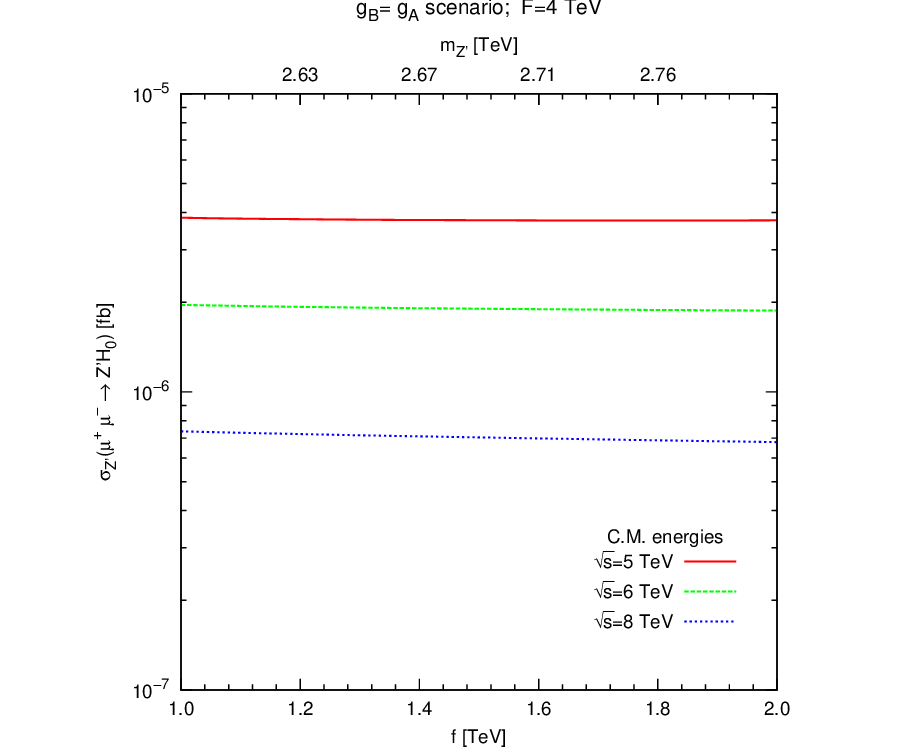}}
\subfloat[]{\includegraphics[width=8.0cm]{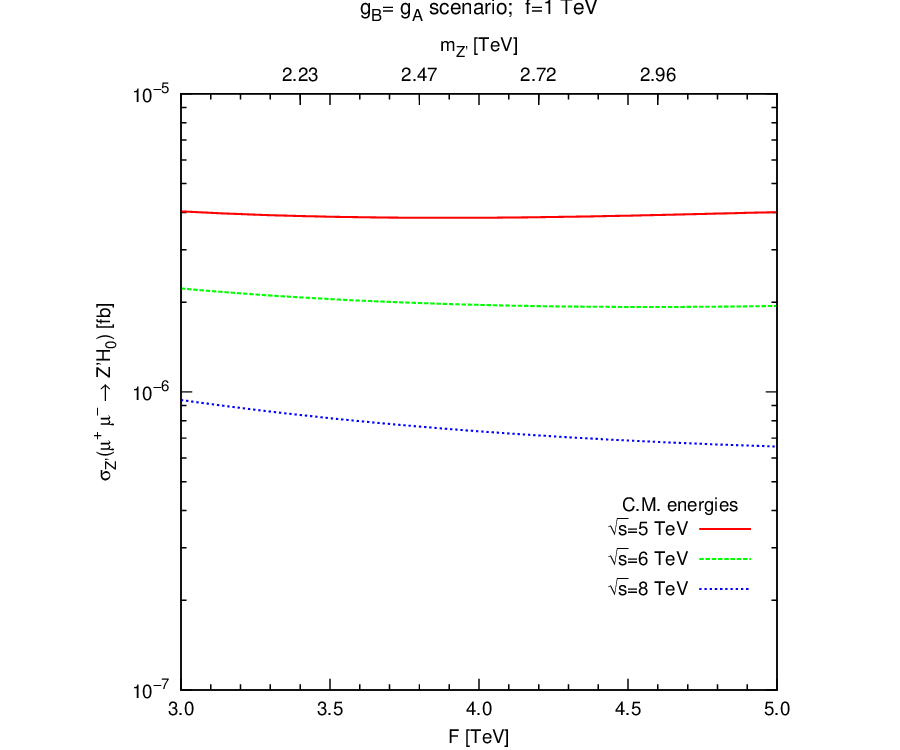}}
        \caption{ a) Cross section of the process $\mu^{+}\mu^{-}\to  Z' \to Z' H_0$   as a function of the $f$ energy scale.
        b) Cross section of the process $\mu^{+}\mu^{-}\to Z' \to Z' H_0$   as a function of the $F$ energy scale.
        The curves are generated for different values of the center-of-mass energy $\sqrt{s}$: $\sqrt{s}=5000$ GeV, $\sqrt{s}=6000$ GeV, and $\sqrt{s}=8000$ GeV.}
\label{PSH0-fF-ap}
\end{figure}

In Fig.~\ref{PSH0-fF-ap}, we show the behavior of the $Z'H_0$ production cross section as a function of the scale $f$ or $F$. The curves shown have been generated for center-of-mass energies of 5000, 6000, and 8000 GeV.  Thus, in the left plot of Figure Fig.~\ref{PSH0-fF-ap}(a), we have $ \sigma_{Z'}(\mu^{+}\mu^{-} \to Z' H_0)$ vs. $f$ (with $F=4000$ GeV). In this scenario, we find that the dominant and subdominant contributions are reached for $\sqrt{s}=5000$ GeV and $\sqrt{s}=6000$ GeV, the numerical values for the $ \sigma_{Z'}(\mu^{+}\mu^{-} \to Z' H_0)$ cross section are $ \sigma_{Z'}(\mu^{+}\mu^{-} \to Z' H_0)=[3.84, 3.76] \times 10^{-6}$ fb and $ \sigma_{Z'}(\mu^{+}\mu^{-} \to Z' H_0)=[1.96, 1.87] \times 10^{-6}$ fb, respectively. 
Complementarily, in the right plot of Fig.~\ref{PSH0-fF-ap}(b), we study  $ \sigma_{Z'}(\mu^{+}\mu^{-} \to Z' H_0)$ vs. $F$ (with $f=1000$ GeV). In this case, the most important contributions are obtained again when $\sqrt{s}=5000$ GeV and $\sqrt{s}=6000$ GeV, such numerical contributions for $ \sigma_{Z'}(\mu^{+}\mu^{-} \to Z' H_0)$ are of the same order of magnitude, namely $ \sigma_{Z'}(\mu^{+}\mu^{-} \to Z' H_0)=[4.03,4.01]\times 10^{-6}$ fb and $ \sigma_{Z'}(\mu^{+}\mu^{-} \to Z' H_0)=[2.22, 1.94 ] \times 10^{-6}$ fb.
 On the upper horizontal axis of the above plots, we also show the gauge boson mass $Z'$ values as the scale of the new physics acquires values in the range of allowed values. The parameters $m_{Z'}$ and $ \sigma_{Z'}(\mu^{+}\mu^{-} \to Z' H_0)$ are sensitive to changes in the values of the new physics scales. The $ \sigma_{Z'}(\mu^{+}\mu^{-} \to Z' H_0)$ cross section takes smaller values for larger values of the $\sqrt{s}$ parameter.

\begin{figure}[H]  
\center
{\includegraphics[width=8.0cm]{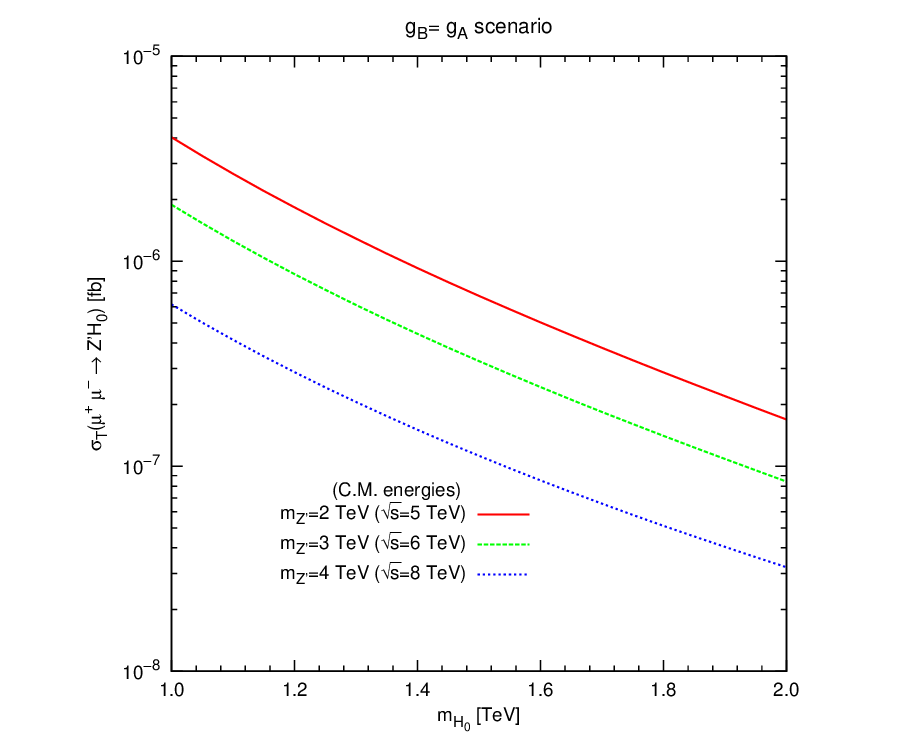}}
\caption{Cross section of the process $\mu^{+}\mu^{-}\to  Z' \to Z' H_0$  as a function of the Higgs boson mass $H_0$. 
 The curves are generated for different values of the center-of-mass energy $\sqrt{s}$: $\sqrt{s}=5000$ GeV, $\sqrt{s}=6000$ GeV, and $\sqrt{s}=8000$ GeV.} 
 \label{PSH0mass-ap}    
\end{figure}

For the present work, the mass of the new heavy Higgs boson $H_0$ has been set around 1000 GeV. However, it would be interesting to analyze the possible dependence of $ \sigma_{Z'}(\mu^{+}\mu^{-} \to Z' H_0)$ on the parameters $m_{H_0}$ and $m_{Z'}$. Therefore, in Fig.~\ref{PSH0mass-ap} we plot $ \sigma_{Z'}(\mu^{+}\mu^{-} \to Z' H_0)$ while varying $m_{H_0}$ in the interval from 1000 to 2000 GeV.  The different curves generated have been obtained for $m_{Z'}=2000$, 3000 and 4000 GeV, where the numerical contributions obtained for $ \sigma_{Z'}(\mu^{+}\mu^{-} \to Z' H_0)$ are the following: $ \sigma_{Z'}(\mu^{+}\mu^{-} \to Z' H_0)=[4.03 \times 10^{-6}, 1.69 \times 10^{-7}]$ fb, $ \sigma_{Z'}(\mu^{+}\mu^{-} \to Z' H_0)=[1.89 \times 10^{-6}, 8.42 \times 10^{-8}]$ fb,  and $ \sigma_{Z'}(\mu^{+}\mu^{-} \to Z' H_0)=[6.16 \times 10^{-7}, 3.22 \times 10^{-8}]$ fb. 
From our numerical analysis, it is clear that the production cross-section of $Z'H_0$ decreases by one or two orders of magnitude as $m_{H_0}$ takes values up to 2000 GeV.  This indicates that  $ \sigma_{Z'}(\mu^{+}\mu^{-} \to Z' H_0)$ depends strongly on the parameters $m_{H_0}$ and $m_{Z'}$.
The $ \sigma_{Z'}(\mu^{+}\mu^{-} \to Z' H_0)$ cross-section also becomes smaller as the $\sqrt{s}$ parameter  increases. 

The production cross-section of $Z'H_0$ provides instead suppressed contributions, which would make it difficult to observe processes associated with $\mu^{+} \mu^{-} \to Z'H_0 $ in a future muon collider. The integrated luminosities and center-of-mass energies of the first stages of the machine would not be sufficient to produce $Z'H_0$ (see Table I of  Ref.~\cite{Accettura:2023ked}).

\section{Conclusions} \label{conclusions}

This paper presents a study of the contributions of the BLHM to the production cross sections of the Higgs-strahlung processes  $\mu^+\mu^- \to Z'h_0$ and $\mu^+\mu^- \to Z'H_0$. The BLHM offers two study scenarios, which arise when the gauge couplings $g_B$ and $g_A$ satisfy the following conditions: $g_B=\frac{1}{2} g_A$ and $g_B=g_A$. In these two scenarios, we explore all the phenomenological implications of the fundamental model parameters on the total production cross-section of $Z'h_0$ and $Z'H_0$. 
Our numerical results show that the contributions of the BLHM to the $Z'h_0$ production cross section are significant since it obtains values of the order of $10^{-1}-10^{-3}$ fb.  In contrast, the values obtained by the $Z'H_0$ cross-section are very small, of the order of $10^{-5}-10^{-8}$ fb.  Notably, the values obtained for the $Z'h_0$ or $Z'H_0$ production cross-sections across different study scenarios exhibit minimal numerical discrepancies.

In order to generate a benchmark in the production of $h_0$ or $H_0$ Higgs bosons in association with a new heavy gauge boson  $Z'$ at a future muon collider.  In this study, we consider integrated luminosities of ${\cal L}=2000, 20 000$ ${\rm fb^{-1}}$ and respective center-of-mass energies of $\sqrt{s}=3000, 10 000$ GeV projected for the initial operational stages of a future muon collider. Tables~\ref{zph13}-\ref{zph14} (when $g_B=\frac{1}{2} g_A$) and Tables~\ref{zph13-apendixx}-\ref{zph14-apendixx} (when $g_B= g_A$) present an estimation of the expected number of $Z'h_0$ events at a muon collider. Concerning $Z'H_0$ production, the cross-section is tiny, necessitating higher luminosity and machine energy for observation.

In addition, we provide Feynman rules for the interaction vertices: scalar-gauge bosons, scalar-fermions (heavy and light), and Higgs boson self-couplings. Thus, the Feynman rules involving scalars are summarized in Tables~\ref{scalar-gauge1}-\ref{3higgs}. 
 It is worth mentioning that among the Higgs self-couplings, the measurement of the triple coupling of the Higgs boson~\cite{Gutierrez-Rodriguez:2011khl,Gutierrez-Rodriguez:2009cak,Gutierrez-Rodriguez:2008evm,Baez:2006dj,Gutierrez-Rodriguez:2005dbe,Gutierrez-Rodriguez:2003tvb} is one of the most important goals of Higgs physics at current and future colliders. The measurement of these triple couplings would provide the first direct information on the Higgs potential responsible for the electroweak symmetry breaking. 
 In this sense, the Feynman rules generated in this work are essential for studying the phenomenology of the scalar bosons of the BLHM. Our results presented in this work can be useful for the scientific community of High-Energy Physics.

\vspace{2cm}

\begin{center}
{\bf Acknowledgements}
\end{center}

 E. Cruz-Albaro appreciates the postdoctoral stay at the Universidad Autnoma de Zacatecas. A. G. R. and D. E. G. thank Sistema Nacional de Investigadoras e Investigadores (Mexico).
 

\vspace{1cm}

\begin{center}
   {\bf Declarations}
\end{center}

 Data Availability Statement: All data generated or analyzed during this study are included in this article.


\newpage

\appendix

\section{Feynman rules of scalars in the BLHM} \label{Rules}

This Appendix provides Feynman rules for interaction vertices: scalars-gauge bosons, scalar-fermions (heavy and light), and Higgs boson self-couplings.

\subsection{Methods}

Feynman rules for interaction vertices are indispensable tools for approaching perturbative calculations in high-energy physics. Their importance lies in the fact that each Feynman diagram that arises can be converted into an algebraic expression (e.g., tensor amplitude), which in turn leads to obtaining a certain particular observable.
In this way, to facilitate the phenomenological study of Higgs boson physics in the BLHM scenario, in this Appendix we provide all the Feynman rules of the interaction vertices obtained in the unitary gauge. These vertices refer to the couplings between scalars-gauge bosons, scalar-fermions, and Higgs boson self-interactions. The complete set of Feynman rules presented in this study was determined using perturbation theory and expanded up to $\mathcal{O}(\frac{1}{f^{2}})$.

The determination of Feynman rules is presented in this section and is carried out using FeynCalc~\cite{Mertig:1990an,Shtabovenko:2020gxv,Shtabovenko:2023idz}, this is an open-source Mathematica package that can be used standalone or integrated into a custom computational setup.
The Mathematica notebooks needed to generate the Feynman rules can be found online at  http://ricaxcan.uaz.edu.mx/jspui/simple-search. 
   For more information on the FeynCalc (FC) computer codes, the interested reader is referred to the FeynCalc manuals in their two versions (https://feyncalc.github.io/): FC guide (stable) and FC guide (development).

\subsection{Feynman rules for scalars-gauge bosons interaction vertices} \label{SB}

In this Subsection, we generate the Feynman rules for all scalars-gauge bosons interactions. 
 These interaction vertices arise from the scalar kinetic term in the Lagrangian given in
Eq.~(\ref{Lcinetico}). From this equation, the exponential functions describing the nonlinear sigma models ($ \Sigma, \Delta $) are expanded as a power series, while the  $A^{a}_{1\, \mu} $, $ A^{a}_{2\, \mu} $  and $ B^{3}_{\mu} $ fields (see Eqs.~(\ref{A1mu})-(\ref{B3mu})) are expressed in terms of the mass eigenstates that are obtained by the authors of Refs.~\cite{Martin:2012kqb,PhenomenologyBLH}. In this way, the Feynman rules corresponding to the interactions between scalars-gauge bosons in the BLHM are given in Tables~\ref{scalar-gauge1} and~\ref{scalar-gauge2}.

\begin{eqnarray}
  A^{1}_{1\mu}&=& \frac{1}{\sqrt{2}} \bigg(  \Big(\text{cos}\, \theta_g\, \text{cos}\,\theta''-  \frac{ \text{cos}\, \theta_W \, \text{sin}\, \theta_g x_s v^{2} }{f^{2}+F^{2}}    \Big)(W^{-}_\mu+W^{+}_\mu ) \nonumber  \\
  &+& \Big(\text{sin}\, \theta_g\, \text{cos}\,\theta''+ \frac{ \text{cos}\, \theta_g \, \text{cos}\, \theta_W x_s v^{2} }{f^{2}+F^{2}}    \Big)(W'^{-}_\mu + W'^{+}_\mu ) \bigg), \label{A1mu} \\
 A^{2}_{1\mu} &=& \frac{i}{\sqrt{2}} \bigg(  \Big(\text{cos}\, \theta_g\, \text{cos}\,\theta''-  \frac{ \text{cos}\, \theta_W \, \text{sin}\, \theta_g x_s v^{2} }{f^{2}+F^{2}}    \Big)(-W^{-}_\mu+W^{+}_\mu ) \nonumber \\
  &+& \Big(\text{sin}\, \theta_g\, \text{cos}\,\theta''+ \frac{ \text{cos}\, \theta_g \, \text{cos}\, \theta_W x_s v^{2} }{f^{2}+F^{2}}    \Big)(-W'^{-}_\mu + W'^{+}_\mu ) \bigg),\\
      A^{3}_{1\mu} &=& \, \text{cos}\, \theta_g \,\text{sin}\, \theta_W\  A_\mu+ \Big(\text{cos}\, \theta_g\, \text{cos}\, \theta_W\, \text{cos}\,\theta'-  \frac{ \text{sin}\, \theta_g \, x_s v^{2} }{f^{2}+F^{2}}    \Big)Z_{\mu} \nonumber \\
   &+& \Big(\text{sin}\, \theta_g\, \text{cos}\, \theta'\, +  \frac{ \text{cos}\, \theta_g \, \text{cos}\, \theta_W\,  x_s v^{2} }{f^{2}+F^{2}}    \Big)Z'_{\mu}, \\
   A^{1}_{2 \mu} &=& \frac{1}{\sqrt{2}} \bigg(  \Big(\text{sin}\, \theta_g\, \text{cos}\,\theta''+  \frac{ \text{cos}\, \theta_W \, \text{cos}\, \theta_g x_s v^{2} }{f^{2}+F^{2}}    \Big)(W^{-}_\mu+W^{+}_\mu ) \nonumber \\
  &-& \Big(\text{cos}\, \theta_g\, \text{cos}\,\theta'' - \frac{ \text{sin}\, \theta_g \, \text{cos}\, \theta_W x_s v^{2} }{f^{2}+F^{2}}    \Big)(W'^{-}_\mu + W'^{+}_\mu ) \bigg),\\
  A^{2}_{2 \mu} &=& \frac{i}{\sqrt{2}} \bigg(  \Big(\text{sin}\, \theta_g\, \text{cos}\,\theta'' +  \frac{ \text{cos}\, \theta_W \, \text{cos}\, \theta_g x_s v^{2} }{f^{2}+F^{2}}    \Big)(-W^{-}_\mu+W^{+}_\mu ) \nonumber \\
  &-& \Big(\text{cos}\, \theta_g\, \text{cos}\,\theta''- \frac{ \text{sin}\, \theta_g \, \text{cos}\, \theta_W x_s v^{2} }{f^{2}+F^{2}}    \Big)(-W'^{-}_\mu + W'^{+}_\mu ) \bigg),
\end{eqnarray}

\begin{eqnarray}
     A^{3}_{2\mu} &=& \, \text{sin}\, \theta_g \,\text{sin}\, \theta_W\  A_\mu+ \Big(\text{sin}\, \theta_g\, \text{cos}\, \theta_W\, \text{cos}\,\theta'+  \frac{ \text{cos}\, \theta_g \, x_s v^{2} }{f^{2}+F^{2}}    \Big)Z_{\mu} \nonumber \\
   &+& \Big(\text{cos}\, \theta_g\, \text{cos}\, \theta'\, - \frac{ \text{sin}\, \theta_g \, \text{cos}\, \theta_W\,  x_s v^{2} }{f^{2}+F^{2}}    \Big)Z'_{\mu}, \\
    B_{3\mu}&=& \, \text{cos}\, \theta_W\  A_\mu - \text{sin}\, \theta_W\, \text{cos}\, \theta' Z_{\mu}- \frac{ \text{sin}\, \theta_W\,  x_s v^{2} }{f^{2}+F^{2}}   Z'_{\mu}, \label{B3mu}
\end{eqnarray}
where 
\begin{eqnarray}
\cos \theta' &=& 1-\frac{1}{2} \frac{v^{2} x_s}{f^{2}+ F^{2}}, \\
\cos \theta'' &=& 1-\frac{1}{2} \frac{v^{2} x_s }{f^{2}+ F^{2}} \cos \theta_W,\\
 x_s &= & \frac{1}{2\, c_W} s_g c_g (s^2_{g} - c^2_{g}).
\end{eqnarray}

\begin{table}[H]
\caption{Feynman rules between scalars-gauge bosons in the BLHM.
\label{scalar-gauge1}}
\begin{tabular}{|p{2.4cm}  p{13.6cm}|}
\hline
\hline
\textbf{Particle} &    \hspace{3.5cm} \textbf{Couplings} \\
\hline
\hline
$h_{0} Z_{\mu} Z_{\nu}  $  &  $ \biggl( \frac{g m_{W}
   \sin (\alpha +\beta )}{c^{2}_{W}}    -\frac{s_{W}^2 v^3
   \left(g^2+g'^2\right)^2 \sin (\alpha +\beta )}{6
   g'^2 f^2} \hfill \break -\frac{s_{W} v^3 x_{s} \left(g^2+g'^2\right) \sin (\alpha
   +\beta ) \left( - c_{g}^2 g g'+c_{g} s_{g}
   s_{W} \left(g^2+g'^2\right)+g g'
   s_{g}^2\right)}{2 c_{g} s_{g} g'^2
   \left(f^2+F^2\right)}  \biggl) g_{\mu \nu}$ \\
\hline
\hline
$h_{0}  Z_{\mu} Z'_{\nu}  $ &   $ \biggl( -\frac{g
   s_{W} v \left(c_{g}^2-s_{g}^2\right)
   \left(g^2+g'^2\right) \sin (\alpha +\beta )}{2 c_{g}
    s_{g} g' }+  \frac{g s_{W} v^3
   \left(c_{g}^2-s_{g}^2\right) \left(g^2+g'^2\right)
   \sin (\alpha +\beta )}{6 c_{g} s_{g} g' f^2 }  \hfill \break  + \frac{v^3
   x_{s} \sin (\alpha +\beta ) \left(c_{g}^2 g g'
   s_{W} \left(g^2+g'^2\right)+2 c_{g} s_{g}
   \left(g^4 s_{W}^2+g^2 g'^2 \left(2
   s_{W}^2+1\right)+g'^4 s_{W}^2\right)-g g'
   s_{g}^2 s_{W} \left(g^2+g'^2\right)\right)}{2
   c_{g}  s_{g}g'^2 \left(f^2+F^2\right)}\biggl) g_{\mu \nu} $ \\
\hline
\hline
$h_{0}  Z'_{\mu} Z'_{\nu}  $ &   $  \biggl(  -\frac{1}{2} g^2 v \sin (\alpha +\beta )  -\frac{g^2 v \sin (\alpha
   +\beta ) \left(c_{g}^4 v^2+s_{g}^4 v^2\right)}{12
   c_{g}^2 s_{g}^2 f^2 }  \hfill \break  -\frac{g v^3 x_{s} \sin (\alpha +\beta ) \left(c_{g}^2
   s_{W} \left(g^2+g'^2\right)-c_{g} g g'
   s_{g}-s_{g}^2 s_{W}
   \left(g^2+g'^2\right)\right)}{2 c_{g} s_{g} g'
   \left(f^2+F^2\right)} \biggl) g_{\mu \nu}$ \\
\hline
\hline
$H_{0} Z_{\mu} Z_{\nu}  $   & $  \biggl( \frac{s_{W}^2 v
   \left(g^2+g'^2\right)^2 \cos (\alpha +\beta )}{2 g'^2}  - \frac{s_{W}^2 v^3
   \left(g^2+g'^2\right)^2 \cos (\alpha +\beta )}{6
   g'^2 f^2 } \hfill \break -\frac{s_{W} v^3 x_{s} \left(g^2+g'^2\right) \cos
   (\alpha +\beta ) \left(c_{g}^2 (-g) g'+c_{g}
   s_{g} s_{W} \left(g^2+g'^2\right)+g g'
   s_{g}^2\right)}{2 c_{g}  s_{g}g'^2
   \left(f^2+F^2\right)}  \biggl) g_{\mu \nu}$ \\
\hline
\hline
$ H_{0} Z_{\mu} Z'_{\nu}  $   & $\biggl( -\frac{g
   s_{W} v \left(c_{g}^2-s_{g}^2\right)
   \left(g^2+g'^2\right) \cos (\alpha +\beta )}{2 c_{g}
    s_{g} g' }+\frac{g s_{W} v^3
   \left(c_{g}^2-s_{g}^2\right) \left(g^2+g'^2\right)
   \cos (\alpha +\beta )}{6 c_{g} s_{g} g' f^2  } \hfill \break + \frac{v^3
   x_{s} \cos (\alpha +\beta ) \left(c_{g}^2 g g'
   s_{W} \left(g^2+g'^2\right)+2 c_{g} s_{g}
   \left(g^4 s_{W}^2+g^2 g'^2 \left(2
   s_{W}^2+1\right)+g'^4 s_{W}^2\right)-g g'
   s_{g}^2 s_{W} \left(g^2+g'^2\right)\right)}{2
   c_{g}  s_{g} g'^2 \left(f^2+F^2\right)} \biggl) g_{\mu \nu}$ \\
\hline
\hline
$H_{0} Z'_{\mu} Z'_{\nu} $    & $ \biggl(  -\frac{1}{2} g^2 v \cos (\alpha +\beta ) - \frac{g^2 v \cos (\alpha
   +\beta ) \left(c_{g}^4 v^2+s_{g}^4 v^2\right)}{12
   c_{g}^2 s_{g}^2 f^2} \hfill \break -\frac{g v^3 x_{s} \cos (\alpha +\beta ) \left(c_{g}^2
   s_{W} \left(g^2+g'^2\right)-c_{g} g g'
   s_{g}-s_{g}^2 s_{W}
   \left(g^2+g'^2\right)\right)}{2 c_{g}  s_{g} g'
   \left(f^2+F^2\right)} \biggl) g_{\mu \nu} $ \\
\hline
\hline
$h_{0}  W_{\mu} W_{\nu} $   & $ \biggl( g m_{W} \sin (\alpha +\beta ) -\frac{4 m_{W}^3
   \sin (\alpha +\beta )}{3 g f^2} + \frac{4
   m_{W}^3 s_{W} x_{s} \left(c_{g}^2-c_{g}
   s_{g}-s_{g}^2\right) \sin (\alpha +\beta )}{c_{g}
    s_{g} g' \left(f^2+F^2\right)}  \biggl) g_{\mu \nu}$ \\
\hline
\hline
$h_{0}  W_{\mu} W'_{\nu}$    & $ \biggl( \ -\frac{g^2 v
   \left(c_{g}^2-s_{g}^2\right) \sin (\alpha +\beta )}{2
   c_{g} s_{g}} + \frac{g^2 v^3
   \left(c_{g}^2-s_{g}^2\right) \sin (\alpha +\beta )}{6
   c_{g} s_{g} f^2}  +\frac{g^3 s_{W}
   v^3 x_{s} \left(c_{g}^2+4 c_{g}
   s_{g}-s_{g}^2\right) \sin (\alpha  +\beta )}{2 c_{g}
    s_{g} g' \left(f^2+F^2\right)} \biggl) g_{\mu \nu} $ \\
\hline
\hline
$h_{0}  W'_{\mu} W'_{\nu}  $    & $ \biggl(  \ -\frac{1}{2} g^2 v
   \sin (\alpha +\beta )  -\frac{g^2 v^3
   \left(c_{g}^4+s_{g}^4\right) \sin (\alpha +\beta )}{12
   c_{g}^2 s_{g}^2 f^2} -\frac{g^3 s_{W}
   v^3 x_{s} \left(c_{g}^2-c_{g}
   s_{g}-s_{g}^2\right) \sin (\alpha +\beta )}{2 c_{g}
    s_{g} g' \left(f^2+F^2\right)} \biggl) g_{\mu \nu}$ \\
\hline
\hline
$H_{0}  W_{\mu} W_{\nu} $    & $\biggl( \ \frac{1}{2} g^2 v \cos (\alpha +\beta )  - \frac{ g^2 v^3 \cos (\alpha +\beta
   )}{6 f^2}  +  \frac{g^3 s_{W}
   v^3 x_{s} \left(c_{g}^2-c_{g}
   s_{g}-s_{g}^2\right) \cos (\alpha +\beta )}{2 c_{g}
    s_{g} g' \left(f^2+F^2\right)}  \biggl) g_{\mu \nu} $ \\
\hline
\hline
$H_{0}  W_{\mu} W'_{\nu}  $  & $\biggl( \  -\frac{g^2 v
   \left(c_{g}^2-s_{g}^2\right) \cos (\alpha +\beta )}{2
   c_{g} s_{g}}  +\frac{g^2 v^3
   \left(c_{g}^2-s_{g}^2\right) \cos (\alpha +\beta )}{6
   c_{g} s_{g} f^2}  +\frac{g^3 s_{W}
   v^3 x_{s} \left(c_{g}^2+4 c_{g}
   s_{g}-s_{g}^2\right) \cos (\alpha +\beta )}{2 c_{g}
    s_{g} g' \left(f^2+F^2\right)}  \biggl) g_{\mu \nu} $ \\
\hline
\hline
$H_{0}  W'_{\mu} W'_{\nu} $    & $\biggl(  \ \ -\frac{1}{2} g^2 v
   \cos (\alpha +\beta )  -\frac{g^2 v^3
   \left(c_{g}^4+s_{g}^4\right) \cos (\alpha +\beta )}{12
   c_{g}^2 s_{g}^2 f^2}  -\frac{g^3 s_{W}
   v^3 x_{s} \left(c_{g}^2-c_{g}
   s_{g}-s_{g}^2\right) \cos (\alpha +\beta )}{2 c_{g}
    s_{g} g' \left(f^2+F^2\right)}  \biggl) g_{\mu \nu} $ \\
\hline
\hline
\end{tabular}
\end{table}

\begin{table}[H]
\caption{Continuation of Table~\ref{scalar-gauge1}.
\label{scalar-gauge2}}
\begin{tabular}{|p{2.9cm}  p{13.1cm}|}
\hline
\hline
\textbf{Particle} &    \hspace{3.5cm} \textbf{Couplings} \\
\hline
\hline
$h_{0} h_{0} Z_{\mu} Z_{\nu}  $  &  $ \biggl( \frac{ g^2 }{2\, c_{W}^2}
   +\frac{ v^2 g^2 (\cos (2 (\alpha +\beta ))-2)}{6 c_{W}^2\, f^2} 
-\frac{ v^2 g^2 x_{s} \left( c_{W}\, (s_{g}^2 - c_{g}^2 )+c_{g} s_{g} \right)}{2
   c_{g} s_{g} \, c_{W}^2 \left(f^2+F^2\right)} 
  \biggl) g_{\mu \nu}$ \\
\hline
\hline
$h_{0} h_{0} Z_{\mu} Z'_{\nu}  $ &   $ \biggl( -\frac{g^2
   \left(c_{g}^2-s_{g}^2\right)}{2 c_{g} c_{W}
   s_{g}} - \frac{g^2 v^2
   \left(c_{g}^2-s_{g}^2\right) (\cos (2 (\alpha +\beta
   ))-2)}{6 c_{g} c_{W} s_{g} f^2} \hfill \break + \frac{g^2 v^2 x_{s} \left(c_{g}^2 c_{W}+2 c_{g}
   s_{g} \left(c_{W}^4+c_{W}^2 \left(2
   s_{W}^2+1\right)+s_{W}^4\right)-c_{W}
   s_{g}^2\right)}{2 c_{g} c_{W}^2 s_{g}
   \left(f^2+F^2\right)} 
  \biggl) g_{\mu \nu} $ \\
\hline
\hline
$h_{0} h_{0}  Z'_{\mu} Z'_{\nu}  $ &   $  \biggl(  -\frac{g^2}{2} + \frac{g^2 v^2
   \left(c_{g}^4+s_{g}^4\right) (\cos (2 (\alpha +\beta
   ))-2)}{12 c_{g}^2 s_{g}^2 f^2}  +\frac{g^2 v^2 x_{s}
   \left(-c_{g}^2+c_{g} c_{W}
   s_{g}+s_{g}^2\right)}{2 c_{g} c_{W} s_{g}
   \left(f^2+F^2\right)}   
    \biggl) g_{\mu \nu}$ \\
\hline
\hline
$H_{0} H_{0} Z_{\mu} Z_{\nu}  $   & $  \biggl( \frac{g^2}{2
   c_{W}^2}  - \frac{g^2 v^2 (\cos
   (2 (\alpha +\beta ))+2)}{6 c_{W}^2 f^{2}} -\frac{g^2 v^2 x_{s} \left(c_{g}^2 (-c_{W})+c_{g}
   s_{g}+c_{W} s_{g}^2\right)}{2 c_{g} c_{W}^2
   s_{g} \left(f^2+F^2\right)} 
\biggl) g_{\mu \nu}$ \\
\hline
\hline
$ H_{0} H_{0} Z_{\mu} Z'_{\nu}  $   & $\biggl(   -\frac{g^2
   \left(c_{g}^2-s_{g}^2\right)}{2 c_{g} c_{W}
   s_{g}}  + \frac{g^2 v^2 \left(c_{g}^2-s_{g}^2\right)
   (\cos (2 (\alpha +\beta ))+2)}{6 c_{g} c_{W}
   s_{g} f^2}  \hfill \break +\frac{g^2 v^2 x_{s} \left(c_{g}^2
   c_{W}+2 c_{g} s_{g} \left(c_{W}^4+c_{W}^2
   \left(2 s_{W}^2+1\right)+s_{W}^4\right)-c_{W}
   s_{g}^2\right)}{2 c_{g} c_{W}^2 s_{g}
   \left(f^2+F^2\right)} 
 \biggl) g_{\mu \nu}$ \\
\hline
\hline
$H_{0} H_{0} Z'_{\mu} Z'_{\nu} $    & $ \biggl( -\frac{g^2}{2}  -\frac{g^2 v^2
   \left(c_{g}^4+s_{g}^4\right) (\cos (2 (\alpha +\beta
   ))+2)}{12 c_{g}^2 s_{g}^2 f^2 } +\frac{g^2 v^2 x_{s}
   \left(-c_{g}^2+c_{g} c_{W}
   s_{g}+s_{g}^2\right)}{2 c_{g} c_{W} s_{g}
   \left(f^2+F^2\right)} 
    \biggl) g_{\mu \nu} $ \\
\hline
\hline
$h_0 h_0 W^{+}_{\mu} W^{-}_{\nu}  $  & $ \biggl(  \frac{g^2}{2} +  \frac{ g^2 v^2
   (\cos (2 (\alpha +\beta ))-2)}{6 f^2} +\frac{c_{W} g^2 v^2 x_{s}
   \left(c_{g}^2-c_{g} s_{g}-s_{g}^2\right)}{2
   c_{g} s_{g} \left(f^2+F^2\right)}
  \biggl) g_{\mu \nu} $ \\
\hline
\hline
$h_0 h_0 W^{+}_{\mu}  W^{\prime -}_{\nu}  $ &  $ \biggl(  \frac{g^2
   \left(s_{g}^2-c_{g}^2\right)}{2 c_{g} s_{g}}  -   \frac{g^2 v^2
   \left(c_{g}^2-s_{g}^2\right) (\cos (2 (\alpha +\beta
   ))-2)}{6 c_{g} s_{g} f^2} + \frac{c_{W} g^2 v^2 x_{s} \left(c_{g}^2+4 c_{g}
   s_{g}-s_{g}^2\right)}{2 c_{g} s_{g} 
   \left(f^2+F^2\right)}
  \biggl) g_{\mu \nu} $\\
\hline
\hline
$ h_0 h_0 W^{\prime +}_{\mu} W^{\prime -}_{\nu} $ &  $ \biggl(  -\frac{g^2}{2} +  \frac{g^2
   v^2 \left(c_{g}^4+s_{g}^4\right) (\cos (2 (\alpha +\beta
   ))-2)}{12 c_{g}^2 s_{g}^2 f^{2}}  +\frac{c_{W} g^2 v^2 x_{s}
   \left(-c_{g}^2+c_{g} s_{g}+s_{g}^2\right)}{2
   c_{g} s_{g} \left(f^2+F^2\right)}
  \biggl) g_{\mu \nu} $ \\
\hline
\hline
$ H_0 H_0 W^{+}_{\mu} W^{-}_{\nu}  $   & $ \biggl(  \frac{g^2}{2} - \frac{ g^2 v^2
   (\cos (2 (\alpha +\beta ))+2)}{6 f^2} +  \frac{c_{W} g^2 v^2 x_{s} \left(c_{g}^2-c_{g}
   s_{g}-s_{g}^2\right)}{2 c_{g} s_{g}
   \left(f^2+F^2\right)} 
  \biggl) g_{\mu \nu} $ \\
\hline
\hline
$ H_0 H_0 W^{+}_{\mu}  W^{\prime -}_{\nu} $   & $ \biggl(   \frac{g^2
   \left(s_{g}^2 -c_{g}^2\right)}{2 c_{g} s_{g}} +  \frac{g^2 v^2
   \left(c_{g}^2-s_{g}^2\right) (\cos (2 (\alpha +\beta
   ))+2)}{6 c_{g} s_{g} f^2} + \frac{c_{W} g^2 v^2 x_{s} \left(c_{g}^2+4 c_{g}
   s_{g}-s_{g}^2\right)}{2 c_{g} s_{g}
   \left(f^2+F^2\right)} 
  \biggl) g_{\mu \nu} $ \\
\hline
\hline
$  H_0 H_0  W^{\prime +}_{\mu} W^{\prime -}_{\nu} $    & $ \biggl(   -\frac{g^2}{2}  -\frac{g^2
   v^2 \left(c_{g}^4+s_{g}^4\right) (\cos (2 (\alpha +\beta
   ))+2)}{12 c_{g}^2 s_{g}^2 f^2} +  \frac{c_{W} g^2 v^2 x_{s} \left(-c_{g}^2+c_{g}
   s_{g}+s_{g}^2\right)}{2 c_{g} s_{g}
   \left(f^2+F^2\right)}
  \biggl) g_{\mu \nu} $ \\
\hline
\hline
\end{tabular}
\end{table}

\subsection{Feynman rules for scalar-fermions interaction vertices } \label{SF}

From Eq.~(\ref{Ltop}), it is possible to generate the Feynman rules corresponding to the scalar-fermions interactions in the BLHM, 
these Feynman rules are given in  Refs.~\cite{Cruz-Albaro:2022kty,Cruz-Albaro:2022lks}


\subsection{Feynman rules for Higgs boson self-couplings } \label{S}

The Higgs boson plays an important role in the SM~\cite{SM-1967} because it generates the masses of all the elementary particles (leptons, quarks, and gauge bosons). However, the Higgs-boson sector is the least tested in the SM, particularly the Higgs-boson self-interaction. In addition, measuring $h_0 h_0$ production at various colliders is necessary to verify the Higgs mechanism of electroweak symmetry breaking.
Furthermore, it is worth mentioning that it is not ruled out the possibility of realization of beyond SM models which include both neutral Higgs scalars $h_0, H_0$, charged Higgs scalars $H^\pm$ and pseudoscalares $A_0$ with masses  $m_{h_0,H_0}$, $m_{H^\pm}$ and  $m_{A_0}$, respectively.
In this regard, within the SM-like scenario in the BLHM, these measurements can give information regarding new physics BSM.

The BLHM Higgs fields~\cite{BLHM-2010,Kalyniak:2013eva}, $h_1$ and $h_2$, form a Higgs potential $ V_{Higgs} $ of two Higgs doublets where each of the doublets are written as 4's of $SO(4)$. The components of the doublets are explicitly expressed in Eqs.~(\ref{h11})-(\ref{h24}).
Through the $ V_{Higgs} $ potential (see Eq.~(\ref{Vhiggs})), the self-interactions between the Higgs bosons are generated.


The Feynman rules for Higgs boson self-couplings are summarized in Table~\ref{3higgs}.

\begin{table}[H]
\caption{Feynman rules involving Higgs boson self-couplings in the BLHM.
\label{3higgs}}
\begin{tabular}{|p{5.6cm} p{9.5cm}|}
\hline
\hline
\textbf{Particle} &    \hspace{1.8cm} \textbf{Couplings} \\
\hline
\hline
$h_{0} h_{0} h_{0} $    & $ \frac{1}{2} v \lambda_{0} \cos (\alpha -\beta ) \sin (2 \alpha ) $ \\
\hline
$H_{0} H_{0} H_{0} $   & $ \frac{1}{2} v \lambda_{0} \sin (2 \alpha ) \sin (\alpha -\beta )  $ \\
\hline
$h_{0} h_{0} H_{0} $    & $ \frac{1}{4} v \lambda_{0} (3 \cos (3 \alpha -\beta )+\cos (\alpha
   +\beta ))  $ \\
\hline
$h_{0} H_{0} H_{0} $   & $ \frac{1}{4} v \lambda_{0} (\sin (\alpha +\beta )-3 \sin (3 \alpha
   -\beta ))  $ \\
\hline
$h_{0} A_{0} A_{0} $   & $ -\frac{1}{2} v \lambda_{0} \cos (\alpha -\beta ) \sin (2 \beta)  $ \\
\hline
$ H_{0} A_{0} A_{0} $  & $ \frac{1}{2} v \lambda_{0} \sin (\alpha -\beta ) \sin (2 \beta ) $ \\
\hline
$h_{0} H^{+} H^{-} $   & $  v \lambda_{0} \cos (\alpha -\beta ) \sin (2 \beta ) $ \\
\hline
$H_{0} H^{+} H^{-} $   & $  -v \lambda_{0} \sin (\alpha -\beta ) \sin (2 \beta ) $ \\
\hline
$h_{0} h_{0} h_{0} h_{0}$   & $  \frac{1}{8} \lambda_{0} \sin ^2(2 \alpha ) $ \\
\hline
$H_{0} H_{0} H_{0} H_{0}$   & $  \frac{1}{8} \lambda_{0} \sin ^2(2 \alpha ) $ \\
\hline
$A_{0} A_{0} A_{0} A_{0}$   & $  \frac{1}{8} \lambda_{0} \sin ^2(2 \beta ) $ \\
\hline
$h_{0} H_{0} H_{0} H_{0}$   & $ -\frac{1}{4} \lambda_{0} \sin (4 \alpha ) $ \\
\hline
$h_{0} h_{0} H_{0} H_{0}$   & $ \frac{1}{8} \lambda_{0} (3 \cos (4 \alpha )+1)  $ \\
\hline
$h_{0} h_{0} h_{0} H_{0}$   & $ \frac{1}{4} \lambda_{0} \sin (4 \alpha ) $ \\
\hline
$h_{0} h_{0} A_{0} A_{0}$   & $  -\lambda_{0} \cos (\alpha ) \cos (\beta ) \sin (\alpha ) \sin
   (\beta ) $ \\
\hline
$h_{0} H_{0} A_{0} A_{0}$   & $  -\frac{1}{2} \lambda_{0} \cos (2 \alpha ) \sin (2 \beta ) $ \\
\hline
$H_{0} H_{0} A_{0} A_{0}$   & $  \lambda_{0} \cos (\alpha ) \cos (\beta ) \sin (\alpha ) \sin
   (\beta ) $ \\
\hline
$h_{0} h_{0}H^{+} H^{-} $   & $ 2 \lambda_{0} \cos (\alpha ) \cos (\beta ) \sin (\alpha ) \sin
   (\beta )  $ \\
\hline
$H_{0} H_{0} H^{+} H^{-} $   & $   -2 \lambda_{0} \cos (\alpha ) \cos (\beta ) \sin (\alpha ) \sin
   (\beta ) $ \\
\hline
$h_{0} H_{0} H^{+} H^{-} $   & $ \lambda_{0} \cos (2 \alpha ) \sin (2 \beta )  $ \\
\hline
$A_{0} A_{0} H^{+} H^{-} $   & $  -2 \lambda_{0} \cos ^2(\beta ) \sin ^2(\beta )  $ \\
\hline
$  H^{+} H^{+} H^{-}H^{-} $   & $ 2 \lambda_{0} \cos ^2(\beta ) \sin ^2(\beta )  $ \\
\hline
\end{tabular}
\end{table}


\newpage

\end{document}